\documentclass{iucrjournals}

\usepackage{amsmath,amssymb,amsthm}
\usepackage{booktabs}
\usepackage{algorithm}
\usepackage{algpseudocode}
\nolinenumbers

\newtheorem{theorem}{Theorem}

\newcommand{\FN}{\operatorname{FoldedNormal}}
\newcommand{\N}{\operatorname{Normal}}
\newcommand{\R}{\mathbb{R}}
\newcommand{\snr}{I/\sigma_{I}}

\DeclareMathOperator{\Var}{\mathbb{V}}
\DeclareMathOperator{\E}{\mathbb{E}}

\DeclareMathOperator{\Normal}{Normal}

\DeclareMathOperator{\Rice}{Rice}
\newcommand{\erf}{\operatorname{erf}}

\newcommand{\Iobs}{I_{\text{obs}}}
\newcommand{\prog}[1]{\textit{#1}}
\newcommand{\vx}{\boldsymbol{x}}
\newcommand{\vz}{\boldsymbol{z}}

\newcommand{\vF}{\boldsymbol{F}}
\newcommand{\vh}{\boldsymbol{h}}
\newcommand{\vI}{\boldsymbol{I}}
\newcommand{\vD}{\boldsymbol{D}}
\newcommand{\vzero}{\mathbf{0}}
\newcommand{\veps}{\boldsymbol{\epsilon}}
\newcommand{\vmu}{\boldsymbol{\mu}}
\newcommand{\norm}[1]{\lVert#1\rVert}

\newcommand{\Gimp}{\hat{\mathcal{G}}^{\mathrm{I}}}
\newcommand{\Gimpmu}{\hat{\mathcal{G}}_{\mu}^{\mathrm{I}}}

\newcommand{\Gimpnu}{\hat{\mathcal{G}}_{\nu}^{\mathrm{I}}}
\newcommand{\Gimpsigma}{\hat{\mathcal{G}}_{\sigma}^{\mathrm{I}}}
\newcommand{\Gimptheta}{\hat{\mathcal{G}}_{\theta}^{\mathrm{I}}}

\newcommand{\Gexp}{\hat{\mathcal{G}}^{\mathrm{E}}}
\newcommand{\Gexpmu}{\hat{\mathcal{G}}_{\mu}^{\mathrm{E}}}
\newcommand{\Gexpmuone}{\hat{\mathcal{G}}_{\mu_1}^{\mathrm{E}}}
\newcommand{\Gexpmutwo}{\hat{\mathcal{G}}_{\mu_2}^{\mathrm{E}}}
\newcommand{\Gexpnu}{\hat{\mathcal{G}}_{\nu}^{\mathrm{E}}}
\newcommand{\Gexpsigma}{\hat{\mathcal{G}}_{\sigma}^{\mathrm{E}}}
\newcommand{\Gexptheta}{\hat{\mathcal{G}}_{\theta}^{\mathrm{E}}}

\newcommand{\cdf}{F_Y(Y;\theta)}

\newcommand{\sign}{\operatorname{sign}}

\title{Implicit and explicit reparameterizations of the Folded Normal and Rice distributions}

\author[a,b]{Luis Aldama\IUCrCemaillink{luis_aldama@g.harvard.edu}\IUCrOrcidlink{0000-0002-3211-4112}}%
\author[c]{Kevin Dalton\IUCrEmaillink{kmdalton@slac.stanford.edu}\IUCrOrcidlink{0000-0001-9396-085X}}%
\author[d]{Minhuan Li\IUCrEmaillink{minhuanli@flatironinstitute.org}\IUCrOrcidlink{0000-0003-0560-7533}}%
\author[b]{Doeke Hekstra\IUCrCemaillink{doeke_hekstra@harvard.edu}\IUCrOrcidlink{0000-0003-2332-9223}}%

\affil[a]{The Harvard Biophysics Graduate Program, Harvard University, Cambridge, Massachusetts, United States}
\affil[b]{Department of Molecular and Cellular Biology, and John A. Paulson School of Engineering and Applied Sciences, Harvard University, Cambridge, Massachusetts, United States}
\affil[c]{LCLS Data Systems, SLAC National Accelerator Laboratory, Menlo Park, CA 94025.}
\affil[d]{Center for Computational Mathematics, Flatiron Institute, New York, NY, USA}

\begin{document}
\maketitle

\begin{synopsis}
  Implicit and explicit reparameterized gradient estimators for the folded Normal and Rice distributions are derived, implemented, and applied to French--Wilson scaling of X-ray diffraction data by variational inference.
\end{synopsis}

\begin{abstract}
  Variational inference is a flexible tool for modeling experimental data and has proven useful in crystallography.
  Pathwise gradient estimators provide low-variance gradients for reparameterized variational inference, but are not readily available for some distributions central to crystallography.
  This work derives and implements implicit and explicit reparameterized gradient estimators for the folded Normal and Rice distributions in \prog{PyTorch}, and compares their variance analytically and empirically.
  In each case the implicit estimator achieves lower variance for the location derivative, particularly at small location-to-scale ratios, while the scale-derivative variances are nearly equal.
  For both distributions, implicit estimator is proved to be the conditional expectation of the explicit estimator given the sampled magnitude, so its variance is never larger.
 As a practical application, French-Wilson scaling is formulated as a variational inference problem, using the folded Normal and Rice     distributions as variational families for centric and acentric reflections, respectively.
  Both estimators recover the exact posterior to better than $1\%$ on the amplitude, and converge at the same rate.
  The implementations are available in the open-source package \prog{rs-distributions}.
\end{abstract}


\section{Introduction}
Stochastic gradient estimators are central to learning algorithms with probabilistic objectives.
They are widely used in machine learning, reinforcement learning, and sensitivity analysis, where gradients must be computed through random variables.
Common estimators include the score-function, pathwise, and measure-valued methods \cite{2020mohamed}.
Among these, pathwise estimation, also known as the reparameterization trick, provides low-variance gradient estimates for differentiable probabilistic models \cite{2019kingma,2014rezende,2018sutton,1998jordan,2017blei}.

In this work we implement explicit and implicit pathwise estimators for the folded Normal (henceforth $\FN$) and Rice distributions.
The $\FN$ distribution was first characterized by \citeasnoun{1961leone}, but the underlying model appears in X-ray crystallography as early as 1956, in the heavy-atom phasing analysis of \citeasnoun{1956woolfson}.
Beyond crystallography, the $\FN$ appears in applications ranging from biomedical modeling \cite{2020liu} to economics \cite{2010liao} and signal processing \cite{2019reig}.
The Rice distribution, introduced by \citeasnoun{1944rice}, is the distribution of the magnitude of a complex normal variable with nonzero mean and isotropic covariance. 
It arises in magnetic resonance imaging (MRI) \cite{1995gudbjartsson,2010lauwers}, wireless communications \cite{1960marcum,2005simon}, and crystallography \cite{2016read}.

Our interest is in crystallography, where these distributions first appeared when \citeasnoun{1949wilson} analyzed the statistics of diffraction intensities.
He showed that the centric and acentric structure factor amplitude distributions are, respectively, the half-normal and Rayleigh distributions, which are special cases of the $\FN$ (with $\mu = 0$) and Rice (with $\nu = 0$) that arise in the absence of prior structural information.
More recently, the $\FN$ and $\Rice$ distributions have taken on a new role in gradient-based structure determination and data reduction, where one conditions on partial structural information and then integrates over the unobserved phase, producing $\FN$ or $\Rice$ likelihood functions \cite{1997bricogne,1996pannu}.
For example, ROCKET refines AlphaFold2 predictions against diffraction data by optimizing intensity-based likelihoods built from the $\FN$ (centric) and Rice (acentric) distributions \cite{2026fadini}.
Similarly, the diffusion-based CrystalBoltz guides sampling with a likelihood built from the same two distributions \cite{2026kim}, and Careless uses the half-normal and Rayleigh distributions as priors on structure factor amplitudes \cite{2022dalton,2023aldama,2025hekstra}.
Such applications, which backpropagate through these likelihoods, motivate efficient and differentiable implementations of the underlying $\FN$ and $\Rice$ sampling paths.

We characterize and compare the behavior of implicit and explicit gradient estimators for these distributions, both analytically and empirically.
For both the $\FN$ and the Rice distributions, we prove a variance ordering between the two estimators by recognizing the implicit estimator as the conditional expectation of the explicit estimator given the sampled magnitude and applying the law of total variance.
As a practical application, we use these reparameterizations to formulate French--Wilson scaling \cite{1978french} as a variational inference problem, demonstrating that our distributions can serve as effective variational families for crystallographic structure factor estimation.
Our implementations are provided in our open-source \prog{PyTorch} package \prog{rs-distributions}.\footnote{\url{https://github.com/rs-station/distributions}}

\subsection{Explicit and implicit gradient estimators}

In this section we set up the notation and mathematical framework for stochastic gradient estimation.
We assume an objective function of the form
\begin{align}
  \label{eq:objective}
  \mathcal{L}(\theta) = \E_{p(\vx;\theta)}[f(\vx)],
\end{align}
where $f$ is a test function of variates $\vx$ from a distribution $p(\vx;\theta)$ with distributional parameters $\theta$.
The gradient of equation~\ref{eq:objective} with respect to $\theta$, $\nabla _{\theta}\mathcal{L}(\theta)$, is our estimand of interest.
Both the implicit and explicit reparameterization methods yield estimators for $ \nabla_{\theta}\mathcal{L}(\theta)$ \cite{2019figurnov,2019kingma}.
These methods are known as \emph{pathwise gradient estimators}, and they use the sampling path of $\vx$.
Samples $\vx$ can be generated by either (1) directly sampling from the distribution $p(\vx;\theta)$, or (2) first drawing samples $\boldsymbol\epsilon$ from a simpler \emph{base distribution} $p(\boldsymbol\epsilon)$ and transforming the sample with a deterministic function $\vx = g(\boldsymbol\epsilon;\theta)$, which is referred to as a \emph{sampling path}.
The sampling path and the Law of the Unconscious Statistician \cite{2020grimmett} allow us to rewrite the expectation in equation~\ref{eq:objective} as one over the base distribution
\begin{align}
  \label{eq:reparameterization}
  \E_{p(\vx;\theta)}[f(\vx)] =
  \E_{p(\boldsymbol\epsilon)}[f(g(\boldsymbol\epsilon;\theta))].
\end{align}
The estimand of interest becomes
\begin{equation}
  \begin{aligned}
    \label{eq:reparam_grads}
    \nabla _{\theta}\mathcal{L}(\theta) & = \nabla _{
    \theta}\E_{p(\boldsymbol\epsilon)}[f(g(\boldsymbol\epsilon;\theta))]
    \\
    & = \nabla_{\theta}\int f(g(\boldsymbol\epsilon; \theta))
    p(\boldsymbol\epsilon) d\boldsymbol\epsilon
    \\
    & = \int \nabla_{\theta} f(g(\boldsymbol\epsilon; \theta))
    p(\boldsymbol\epsilon) d\boldsymbol\epsilon
    \\
    & =
    \E_{p(\boldsymbol\epsilon)}[\nabla_{\theta}f(g(\boldsymbol\epsilon;\theta))],
  \end{aligned}
\end{equation}
where we assume the standard regularity conditions --- almost-sure differentiability of the sampling path and an integrable local dominating function --- that justify exchanging differentiation and integration in the third line \cite{2003glasserman,2020mohamed}.
In \emph{explicit reparameterization} \cite{2019figurnov}, base samples are transformed with a sampling path $g$, giving the Monte Carlo estimator
\begin{align}
  \label{eq:mc_estimator}
  \Gexptheta  = \frac{1}{N}\sum_{n =1}^{N}
  \nabla_{\theta}f(g(\boldsymbol\epsilon_n;\theta)).
\end{align}
Under explicit reparameterization, the sampling path $g(\boldsymbol\epsilon;\theta)$ is built from an invertible \emph{standardization} function $s(\vx;\theta)$ that maps the $\theta$-dependent variable $\vx$ to a variate $\boldsymbol\epsilon = s(\vx;\theta)\sim p(\boldsymbol\epsilon)$, where the base distribution $p(\boldsymbol\epsilon)$ does not depend on $\theta$.
For the $\FN$, the base distribution is the standard normal $\epsilon\sim\N(0,1)$, and the standardization is that of the underlying normal $X\sim\N(\mu,\sigma^2)$.
This standardization is $\epsilon=s(x;\mu,\sigma)=(x-\mu)/\sigma$, so that $s(X;\mu,\sigma)=\epsilon\sim\N(0,1)$.
The sampling path is then a composition of the inverse standardization function, $x = \epsilon \sigma + \mu$, and the absolute-value transform, $g(\epsilon; \mu, \sigma) = |x| = |\epsilon \sigma + \mu|$.
We note that explicit reparameterization requires an invertible standardization function; the outer transform (here $|\cdot|$) does \emph{not} need to be invertible.

For some distributions, the standardization function has no closed-form inverse, or that inverse is too expensive to compute.
In this case, one can use \emph{implicit reparameterization} \cite{2019figurnov}, which does not require inverting the standardization function.
Instead, implicit reparameterization differentiates the standardization relation $\boldsymbol\epsilon = s(\vx;\theta)$ directly.
Holding the standardized sample $\boldsymbol\epsilon$ fixed and differentiating with respect to $\theta$ yields the pathwise derivative $\nabla_\theta \vx = -[\nabla_{\vx} s(\vx;\theta)]^{-1}\nabla_\theta s(\vx;\theta)$, which needs only $s$ and its derivatives, not its inverse.
Here, the Monte Carlo estimator takes the form
\begin{equation}
  \begin{aligned}
    \Gimptheta & = \frac{1}{N}\sum_{n =1}^{N}
    \nabla_{\vx}f(\vx) \nabla_{\theta}\vx
    \\
    &= -\frac{1}{N}\sum_{n=1}^N \nabla_{\vx}f(\vx_n)[\nabla_{\vx}s(\vx_n;\theta)]^{-1} \nabla_{\theta} s(\vx_n;\theta),
  \end{aligned}
  \label{eq:mc_estimator_imp}
\end{equation}
where $\Gimptheta$ denotes the implicit Monte Carlo estimator (equation~\ref{eq:mc_estimator_imp}).
Throughout, the subscript on $\hat{\mathcal G}$ denotes the parameter with respect to which the derivative is taken, and the superscripts E and I distinguish the explicit and implicit estimators.

\section{Implementation}

In this section we derive the implicit and explicit gradient estimators of the $\FN$ and $\Rice$ distributions, and we implement them in \prog{PyTorch} \cite{2019paszke}.
To validate our implementations, we use a test function with analytical derivatives that serve as ground truths.
We also derive the theoretical variance of each estimator, and we check our empirical results against it.

\subsection{FoldedNormal distribution}

Let $X \sim \N(\mu,\sigma^2)$ and let $Y = |X|$, so that $Y \sim \FN(\mu,\sigma)$.
The density of $Y$ is a sum of two $\N$ densities,
\begin{align}
  \label{eq:fn_pdf}
  p(y; \mu, \sigma) = \frac{1}{\sigma\sqrt{2\pi}}\left[\exp\left(-\frac{(y - \mu)^{2}}{2\sigma^{2}}\right) + \exp\left(-\frac{(y + \mu)^{2}}{2\sigma^{2}}\right)\right], \quad y \geq 0,
\end{align}
where the two terms are the contributions of the pre-images $X = +y$ and $X = -y$.
The special case $\mu = 0$ recovers the half-normal distribution, which serves as the centric Wilson prior in Section~\ref{sec:french_wilson}.

\subsubsection{FoldedNormal: test function}
\label{subsubsec:fn_test_func}
We validate our implicit and explicit $\FN$ implementations against the test function $f(Y) = Y^2$, which has closed-form derivatives.
The expectation of the test function, $\E[f(Y)] = \E[Y^2]$, follows from the $\FN$ variance identity
\begin{align}
  \label{eq:fn_var}
  \Var[ Y ] = \mu^{2} + \sigma^{2} - (\E[ Y ])^{2},
\end{align}
and the second-moment expansion
\begin{equation}
  \begin{aligned}
    \E[ Y^{2} ] & = \Var[ Y ] + (\E[ Y ])^{2}\\
    & = \mu^2 + \sigma^2.
  \end{aligned}
\end{equation}
Differentiating with respect to $\mu$ and $\sigma$ gives us the ground truth derivatives
\begin{gather}
  \label{eq:ground_truth_fn_derivatives}
  \frac{\partial}{\partial \mu} \E[ Y^{2} ] = 2\mu
  \quad\text{and}\quad
  \frac{\partial}{\partial \sigma} \E[ Y^{2} ] = 2 \sigma.
\end{gather}

\subsubsection{FoldedNormal: explicit estimator}
\label{subsubsec:fn_exp_pathwise_estimator}
A $\FN(\mu,\sigma)$ random variable is the absolute value of a $\N(\mu,\sigma^2)$ random variable \cite{1961leone}.
The standardization function for a $\N(\mu,\sigma^2)$ distribution is $s(x;\mu,\sigma) = (x-\mu)/\sigma$, so that $s(X;\mu,\sigma) = \epsilon \sim \N(0,1)$.
The sampling path for $Y \sim \FN(\mu,\sigma)$ is given by
\begin{equation}
  Y = g(\epsilon;\mu,\sigma) = |X|, \qquad X = \mu + \sigma \epsilon,
\end{equation}
where $X=s^{-1}(\epsilon; \mu, \sigma)$ is obtained from the inverse standardization function (the location--scale transform).
We use explicit reparameterization to construct a Monte Carlo gradient estimator (equation~\ref{eq:mc_estimator}) of the form
\begin{equation}
  \label{eq:explicit_estimator}
  \begin{aligned}
    \Gexptheta  & = \frac{1}{N}\sum_{n=1}^{N} \nabla_{\theta} f(g(\epsilon_n; \mu, \sigma)) \\
    & = \frac{1}{N}\sum_{n=1}^{N} \nabla_{\theta} f(Y_n) \\
    & = \frac{1}{N}\sum_{n=1}^{N}f'(Y_n)\frac{\partial Y_n}{\partial X_n} \nabla_\theta X_n,
  \end{aligned}
\end{equation}
where $N$ is the number of Monte Carlo samples, $f(Y_n) = Y_n^2$ is the test function, and $\theta = (\mu,\sigma)$.
Taking the $\mu$- and $\sigma$-components of equation~\ref{eq:explicit_estimator} yields estimators of the ground-truth derivatives (equation~\ref{eq:ground_truth_fn_derivatives})
\begin{gather}
  \Gexpmu = \frac{1}{N}\sum_{n=1}^{N}  2X_{n}
  \quad\text{and}\quad
  \Gexpsigma = \frac{1}{N}\sum_{n=1}^{N}  \left(2 \mu
  \epsilon_{n} + 2 \sigma \epsilon_{n}^{2}\right).
\end{gather}
We refer to $\Gexpmu$ and $\Gexpsigma$ as the explicit location- and scale-derivative estimators, respectively.
In Section~S1 of the supporting information, we show that these estimators are unbiased, and that their variance is given by
\begin{gather}
  \label{eq:fn_var_gexp}
  \Var[ \Gexpmu ] = \frac{4 \sigma^{2}}{N}
  \quad\text{and}\quad
  \Var[ \Gexpsigma ] = \frac{4\left(\mu^{2} + 2
  \sigma^{2}\right)}{N}.
\end{gather}

\subsubsection{FoldedNormal: implicit estimator}
\label{subsubsec:fn_imp_pathwise_estimator}
When deriving the explicit pathwise estimator, we used the $\N$ distribution standardization function to construct the sampling path.
Alternatively, we can use the $\FN$ cumulative distribution function (CDF), denoted by $F_Y$, as the standardization function \cite{2019figurnov}.
In this case, the standardization function does not admit a closed-form inverse, so we use implicit reparameterization instead.
The implicit Monte Carlo estimator is given by
\begin{equation}
  \label{eq:implicit_estimator}
  \begin{aligned}
    \Gimptheta & = \frac{1}{N}\sum_{n=1}^{N} f'(Y_n)
    \left(-\frac{\nabla_{\theta}F_Y(Y_n;\theta)}{\partial_{Y}F_Y(Y_n;\theta)}\right) \\
    & = \frac{1}{N}\sum_{n=1}^{N} f'(Y_n)
    \left(-\frac{\nabla_{\theta}F_Y(Y_n;\theta)}{p(Y_n;\theta)}\right),
  \end{aligned}
\end{equation}
where in the second line we used the fact that the derivative of the CDF is the probability density function (PDF) $p(Y_n;\theta)$.
Taking the $\mu$- and $\sigma$-components of equation~\ref{eq:implicit_estimator} yields the following Monte Carlo estimators (derived in Section~S2 of the supporting information),
\begin{align}
  \Gimpmu =\frac{1}{N}\sum_{n=1}^{N}
  2Y_{n}\frac{\phi(a_n) - \phi(b_n)}{\phi(a_n) + \phi(b_n)}
\end{align}
and
\begin{align}
  \Gimpsigma = \frac{1}{N} \sum_{n=1}^{N}
  \frac{2Y_{n}}{\sigma} \frac{(Y_{n} - \mu)\phi(a_n) + (Y_{n} +
  \mu)\phi(b_n)}{\phi(a_n) + \phi(b_n)},
\end{align}
where $\phi$ is the PDF of the standard $\N$ distribution, $a_n = (Y_n - \mu)/\sigma$, and $b_n = (Y_n + \mu)/\sigma$.

\subsubsection{FoldedNormal: variance ordering}
\label{sssec:variance_ordering}

The explicit and implicit estimators defined above produce different per-sample gradient estimates.
The explicit estimator depends on the sign of the underlying $\N$ sample $X$ (through $\partial Y/\partial X = \sign(X)$), whereas the implicit estimator depends only on the $\FN$ sample $Y$.

\begin{theorem}[Variance ordering of $\FN$ derivative estimators]
  \label{thm:variance_ordering}
  Let $Y \sim \FN(\mu, \sigma)$ and let $f : [0, \infty) \to \R$ be continuously differentiable with $\E[(f'(Y))^{2}(1 + Y^{2})] < \infty$.
  For $\theta \in \{\mu,\sigma\}$, the single-sample explicit and implicit derivative estimators of $\partial_\theta\E[f(Y)]$ satisfy
  \begin{align}
    \label{eq:rb_identity}
    \Gimptheta \;=\; \E[\Gexptheta \mid  Y].
  \end{align}
  Consequently, the implicit estimator is unbiased, and its variance is bounded above by the variance of the explicit estimator,
  \begin{align}
    \label{eq:variance_ordering}
    \Var[\Gimptheta] \le \Var[\Gexptheta].
  \end{align}
\end{theorem}
The proof of Theorem~\ref{thm:variance_ordering} is given in Appendix~\ref{appendix:var_proof}.
In the proof of Theorem~\ref{thm:variance_ordering}, we decompose the explicit variance by the law of total variance, yielding the following expression
\begin{equation}
  \Var[\Gexptheta] = \Var[\Gimptheta] + \E\big[\Var[\Gexptheta \mid Y]\big],
\end{equation}
which we use to derive the variance of the implicit location- and scale-derivative estimators:
\begin{align}
  \label{eq:theo_var_fn_implicit}
  \Var[\Gimpmu] &= \frac{1}{N}\left[4\sigma^{2} - \E\big[\Var[\Gexpmu \mid Y]\big]\right], &
  \Var[\Gimpsigma] &= \frac{1}{N}\left[4(\mu^{2} + 2\sigma^{2}) - \E\big[\Var[\Gexpsigma \mid Y]\big]\right].
\end{align}
The term $\E\big[\Var[\Gexptheta \mid Y]\big]$ does not have a closed form but can be approximated with numerical quadrature.
Our full derivation of equation~\ref{eq:theo_var_fn_implicit} is in Appendix~\ref{appendix:fn_gap}.

\subsection{Rice distribution}
\label{subsec:gradients_rice_distribution}

Let $\vz = (z_{1}, z_{2}) \in \R^{2}$ be a bivariate $\N$ vector with independent components $z_{1} \sim \N(\mu_{1},\sigma^{2})$ and $z_{2} \sim \N(\mu_{2},\sigma^{2})$.
Let $R = \norm{\vz} = \sqrt{z_{1}^{2} + z_{2}^{2}}$, so that $R \sim \Rice(\nu,\sigma)$ with $\nu = \norm{\vmu}$ and $\vmu = (\mu_{1},\mu_{2})$.
The Rice law depends on $\vmu$ only through its norm $\nu$; the direction of $\vmu$ is not identifiable.
The density of $R$ is
\begin{align}
  \label{eq:rice_pdf}
  p(r; \nu, \sigma) = \frac{r}{\sigma^{2}} \exp\left(-\frac{r^{2} + \nu^{2}}{2\sigma^{2}}\right) I_{0}\left(\frac{r\nu}{\sigma^{2}}\right), \quad r \geq 0,
\end{align}
where $I_{0}$ is the modified Bessel function of the first kind, order zero.
The special case $\nu = 0$ recovers the Rayleigh distribution, which appears as the acentric Wilson prior in Section~\ref{sec:french_wilson}.

\subsubsection{Rice: test function}
\label{subsubsec:rice_test_fn}
We validate the $\Rice$ estimators against the test function $f(R) = R^2$, whose expectation  $\E[f(R)] = \E[R^2]$ has a closed-form gradient.
Expanding the expectation, we get
\begin{equation}
  \begin{aligned}
    \label{eq:sq_rice_mean}
    \E[R^{2}] & = \E[\norm{\vz}^{2}] \\
    &= \E \left[ z_{1}^{2} + z_{2}^{2} \right ] \\
    & = \E[z_{1}^{2}] + \E[z_{2}^{2}] \\
    & = \Var[z_{1}] + (\E[z_{1}])^{2} + \Var[z_{2}] + (\E[z_{2}])^{2} \\
    & = \sigma^{2} + \mu_{1}^{2} + \sigma^{2} + \mu_{2}^{2} \\
    & = 2\sigma^{2} + \norm{\vmu}^{2} \\
    & = 2 \sigma^{2} + \nu^{2}.
  \end{aligned}
\end{equation}
Differentiating with respect to the Rice parameters $\nu$ and $\sigma$ gives us the ground truth derivatives
\begin{gather}
  \frac{\partial}{\partial \nu} \E[R^2] = 2\nu
  \quad\text{and}\quad
  \frac{\partial}{\partial \sigma} \E[R^2] = 4\sigma.
\end{gather}

\subsubsection{Rice: explicit estimator}
\label{sssec:rice_explicit_estimator}
Each component of $\vz$ follows a $\Normal$ distribution, so the sampling path for $R \sim \Rice(\nu,\sigma)$ can be written in terms of two independent standard Normal variables,
\begin{equation}
  R = g(\epsilon_1, \epsilon_2; \mu_1, \mu_2, \sigma)
  = \sqrt{(\mu_1 + \sigma\epsilon_1)^2 + (\mu_2 + \sigma\epsilon_2)^2},
\end{equation}
where $\epsilon_1,\epsilon_2 \overset{\text{i.i.d.}}{\sim} \N(0,1)$.
Since $\nu = \norm{\vmu}$, explicit reparameterization produces derivatives with respect to $\mu_1$ and $\mu_2$, rather than $\nu$.
Using the sampling path and the test function $f(R) = R^2$, we obtain the unbiased derivative estimators (derived in Section~S3 of the supporting information)

\begin{equation}
  \begin{aligned}
    \Gexpmuone & = \frac{1}{N}\sum_{n=1}^{N}
    \left(2\mu_{1}+2 \sigma\epsilon_1\right),
    \\
    \Gexpmutwo & = \frac{1}{N}\sum_{n=1}^{N}
    \left(2\mu_{2}+2 \sigma\epsilon_2\right),
    \\
    \Gexpsigma  & = \frac{1}{N}\sum_{n=1}^{N}
    \left(2\mu_{1}\epsilon_1 + 2 \sigma\epsilon_1^{2} +
    2\mu_{2}\epsilon_2 + 2 \sigma\epsilon_2^{2}\right).
  \end{aligned}
\end{equation}
Because the Rice law depends on $\vmu$ only through $\nu$, we align the mean with the first coordinate without loss of generality, $\vmu = (\nu, 0)$, so the location derivative is the single-component derivative $\Gexpnu = \Gexpmuone$. For a general mean, the $\nu$-derivative is the projection $\Gexpnu = (\mu_1/\nu)\Gexpmuone + (\mu_2/\nu)\Gexpmutwo$, whose continuous limit at $\nu = 0$ is described in Section~S3 of the supporting information.
The variance of these estimators is given by
\begin{gather}
  \label{eq:rice_var_gexp}
  \Var[\Gexpnu] =  \frac{4\sigma^{2}}{N}
  \quad\text{and}\quad
  \Var[\Gexpsigma] = \frac{4(\nu^{2} + 4\sigma^{2})}{N}.
\end{gather}

\subsubsection{Rice: implicit estimator}
\label{sssec:rice_implicit_estimator}

The CDF of the Rice distribution is given by
\begin{equation}
  F_R(r; \nu, \sigma) = 1 -
  Q_1\left(\frac{\nu}{\sigma},\frac{r}{\sigma}\right),
\end{equation}
where $Q_{1}$ is the first-order Marcum $Q$-function \cite{1960marcum,2005simon}.
The partial derivatives of the $\Rice$ CDF are known \cite{1968pratt}, and we use them to derive the implicit reparameterization gradients in Section~S4 of the supporting information. Writing $\kappa = \nu R/\sigma^{2}$ for the argument of the modified Bessel functions ($I_0$ and $I_1$), the implicit pathwise derivatives are
\begin{align}
  \frac{\partial R}{\partial \nu}    & =
  \frac{I_{1}(\kappa)}{I_{0}(\kappa)}, \\
  \frac{\partial R}{\partial \sigma} & =
  \frac{1}{\sigma}\left(
    R - \nu\frac{I_{1}(\kappa)}{I_{0}(\kappa)}
  \right).
\end{align}
For any differentiable test function $f$, the implicit Monte Carlo gradient estimators are
\begin{equation}
  \label{eq:rice_imp_dnu}
  \Gimpnu
  =\frac{1}{N}\sum_{n=1}^N f'(R_n) \frac{I_1(\kappa_n)}{I_0(\kappa_n)},
\end{equation}
and
\begin{equation}
  \label{eq:rice_imp_dsig}
  \Gimpsigma
  =\frac{1}{N}\sum_{n=1}^N f'(R_n)
  \frac{1}{\sigma}\left(
    R_n-\nu \frac{I_1(\kappa_n)}{I_0(\kappa_n)}
  \right),
\end{equation}
where $\kappa_n = \nu R_n/\sigma^{2}$.

\subsubsection{Rice: variance ordering}
\label{sssec:rice_variance_ordering}

As in the $\FN$ case, the explicit and implicit estimators produce different per-sample gradient estimates.
Writing $\hat{\vz} = \vz/R$ and $\hat{\vmu} = \vmu/\nu$ for the unit sample and mean directions, the explicit estimator depends on the cosine of the angle between the sample and the mean direction, $\hat{\vz}\cdot\hat{\vmu}$, whereas the implicit estimator depends only on the magnitude $R$.

\begin{theorem}[Variance ordering of $\Rice$ derivative estimators]
  \label{thm:rice_variance_ordering}
  Let $R \sim \Rice(\nu, \sigma)$ and let $f : [0, \infty) \to \R$ be continuously differentiable with $\E[(f'(R))^{2}(1 + R^{2})] < \infty$.
  For $\theta \in \{\nu, \sigma\}$, the single-sample explicit and implicit derivative estimators of $\partial_\theta\E[f(R)]$ satisfy
  \begin{align}
    \label{eq:rice_rb_identity}
    \Gimptheta \;=\; \E[\Gexptheta \mid R].
  \end{align}
  Consequently, the implicit estimator is unbiased, and its variance is bounded above by the variance of the explicit estimator,
  \begin{align}
    \label{eq:rice_variance_ordering}
    \Var[\Gimptheta] \le \Var[\Gexptheta].
  \end{align}
\end{theorem}
The proof of Theorem~\ref{thm:rice_variance_ordering} is given in Appendix~\ref{appendix:rice_var_proof}.
This is the exact analogue of Theorem~\ref{thm:variance_ordering}, with the magnitude $R$ playing the role of the $\FN$ sample $Y$, and the sampled direction on the circle playing the role of $\sign(X)$.
In both cases the implicit estimator is the Rao--Blackwellization of the explicit one \cite{1996casella,2025lam}, obtained by conditioning on the magnitude and averaging over the direction.

The variance of the implicit location- and scale-derivative estimators is given by
\begin{align}
  \label{eq:theo_var_rice_implicit}
  \Var[\Gimpnu] &= \frac{1}{N}\left[4\sigma^{2} - \E\big[\Var[\Gexpnu \mid R]\big]\right], &
  \Var[\Gimpsigma] &= \frac{1}{N}\left[4(\nu^{2} + 4\sigma^{2}) - \E\big[\Var[\Gexpsigma \mid R]\big]\right],
\end{align}
with $\E[\Var[\Gexptheta \mid R]]$ again evaluated by numerical quadrature (derivation in Appendix~\ref{appendix:rice_gap}).

\section{Application: French--Wilson scaling}
\label{sec:french_wilson}

X-ray crystallography is routinely used for determining the three-dimensional electron density of a molecule from its diffraction data.
Doing so requires determining complex-valued terms called \emph{structure factors}, which we denote by $\vF$.
In practice, data reduction algorithms are used to estimate the \emph{intensity} ($I$) of each structure factor.
The intensity is related to the squared amplitude ($F$) of the structure factors, $I = F^2 = \vF \vF^*$.
Although the intensity is strictly non-negative by definition, data processing algorithms may return negative intensities.
These negative values cannot be used by subsequent algorithms, and they must either be discarded or mapped to non-negative values.

\citeasnoun{1978french} address negative integrated intensities with a Bayesian correction.
The idea is to use Bayes' theorem to obtain the following posterior distribution
\begin{equation}
  p(F \mid \Iobs) \propto p(\Iobs \mid F) \cdot p(F),
  \label{eq:fw_posterior}
\end{equation}
where $F^2$ is the \emph{true} intensity, $\Iobs$ is a measured intensity, $p(F)$ is a Wilson prior, and $p(\Iobs \mid F)$ is a $\N$ likelihood.
The Wilson prior is given by
\begin{align}
  p(F) &=
  \begin{cases}
    \sqrt{\dfrac{2}{\pi\Sigma}} \exp\left(-\dfrac{F^2}{2\Sigma}\right) & \text{centric}, \\[6pt]
    \dfrac{2F}{\Sigma} \exp\left(-\dfrac{F^2}{\Sigma}\right) & \text{acentric},
  \end{cases}
  \label{eq:wilson_prior}
\end{align}
where the centric prior is a half-normal distribution and the acentric prior is a Rayleigh distribution, the zero-location special cases of the $\FN$ and $\Rice$ families whose estimators we derived above.
Thus, the posterior is proportional to the product of a $\N$ likelihood and Wilson prior, and the posterior has positive support.

To demonstrate one use of our reparameterizations, we approximate the French--Wilson posterior using variational inference \cite{1998jordan,2017blei}.
In particular, we use the $\FN$ and $\Rice$ distributions to approximate the centric and acentric posterior distributions, respectively.
Employing these distributions in this way builds on a history of ``inflated-variance'' Rice distributions that combine prior information on structure factor amplitudes with experimental observations \cite{1979green,2016read}.

Let $q(F)$ denote the variational distribution, where $q(F) = \FN(\mu, \sigma)$ for centric reflections and $q(F) = \Rice(\nu, \sigma)$ for acentric reflections.
The objective function used in variational inference is called the Evidence Lower BOund (ELBO), which is given by
\begin{align}
  \label{eq:fw_elbo}
  \mathcal{L}(\theta) = \E_{q(F)}\left[\log p(I_{\text{obs}} \mid F)\right] - \mathrm{KL}\left(q(F) \;\Vert\; p(F)\right),
\end{align}
where $\E_{q(F)}\left[\log p(I_{\text{obs}} \mid F)\right]$ is the expected log-likelihood, $\mathrm{KL}\left(q(F) \;\Vert\; p(F)\right)$ is the Kullback--Leibler divergence, and $\theta = (\mu, \sigma)$ for centrics or $\theta = (\nu , \sigma)$ for acentrics.
No closed form of the Kullback--Leibler divergence exists for either the $\FN$ or the $\Rice$, so we estimate the ELBO itself with the following Monte Carlo estimator
\begin{align}
  \label{eq:fw_elbo_mc}
  \hat{\mathcal{L}}(\theta) = \frac{1}{N} \sum_{n=1}^{N} \left[\log p(I_{\text{obs}} \mid F_n) + \log p(F_n) - \log q(F_n)\right], \qquad F_n \sim q(F).
\end{align}

We evaluate on a synthetic dataset of $2{,}240$ $(I_{\text{obs}}, \sigma_I, \Sigma)$ samples, and on a hen egg-white lysozyme experimental dataset (PDB: 9B7C).
The supporting information gives the optimization protocol and learning rates (Section~S5, Table~S1), and the datasets and the reference calculations they are scored against (Section~S6).

\section{Results}

\subsection{Variance of the FoldedNormal estimators}
\label{subsec:variance_comparison}

\begin{figure}[h!tb]
  \centering
  \includegraphics[width=\textwidth]{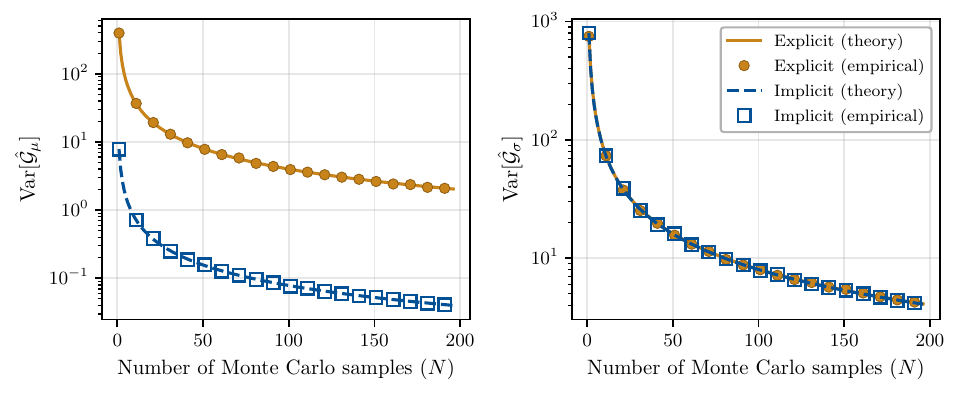}
  \caption{
    Theoretical and empirical variance of the $\FN$ pathwise estimators as a function of Monte Carlo sample size $N$, at $\mu = 1, \sigma = 10$ (small $|\mu|/\sigma$).
    Left, the derivative with respect to $\mu$; right, the derivative with respect to $\sigma$.
    The two empirical curves match their respective theoretical predictions: $\Var[\Gexpmu] = 4\sigma^{2}/N$ for the explicit estimator and equation~\ref{eq:theo_var_fn_implicit} for the implicit estimator (left), where the expected conditional variance is the analytical gap from Theorem~\ref{thm:variance_ordering}.
    For the $\sigma$ derivative (right) both empirical curves coincide with the theoretical prediction of equation~\ref{eq:fn_var_gexp}.
    At this parameter setting, the implicit $\mu$ derivative estimator achieves variance roughly 50-fold smaller than the explicit estimator.
  }
  \label{fig:th_vs_exp_varn}
\end{figure}

\begin{figure}[h!tb]
  \centering
  \includegraphics[width=\textwidth]{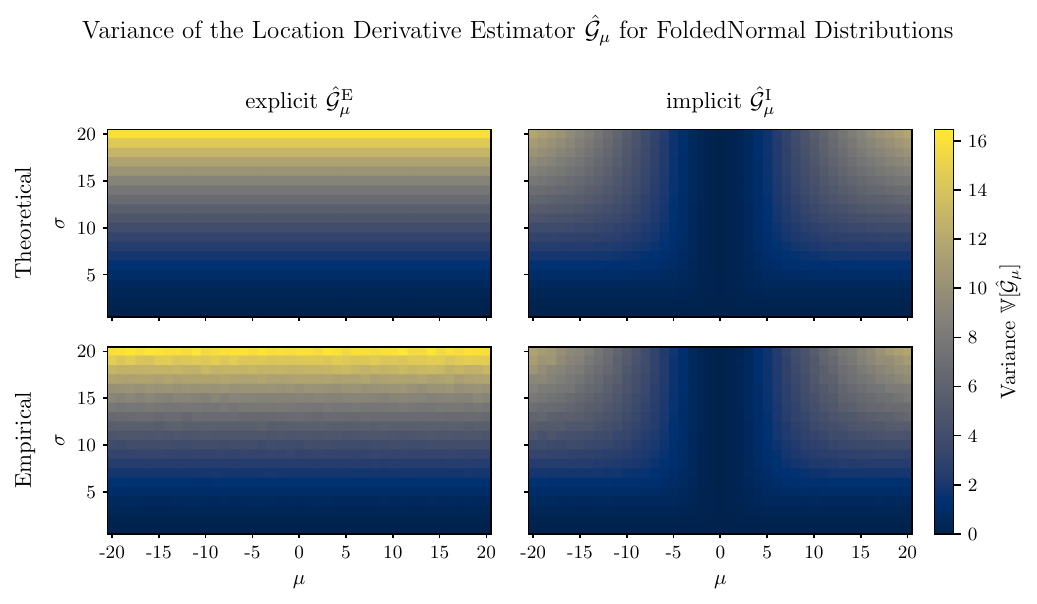}
  \caption{
    Theoretical (top) and empirical (bottom) variance of the explicit ($\Gexpmu$) and implicit ($\Gimpmu$) $\FN$ derivative estimators of $\partial_{\mu}\E[Y^{2}]$ as a function of $\mu$ and $\sigma$.
    The colorbar denotes the variance value for each $(\mu,\sigma)$ pair.
    The empirical variance was computed across $10{,}000$ simulations using $100$ Monte Carlo samples each.
    The theoretical variance was calculated using equations~\ref{eq:fn_var_gexp} and \ref{eq:theo_var_fn_implicit}, for the explicit and implicit estimators, respectively.
  }
  \label{fig:fn_imp_exp_dfdmu_var_subplots}
\end{figure}

\begin{figure}[h!tb]
  \centering
  \includegraphics{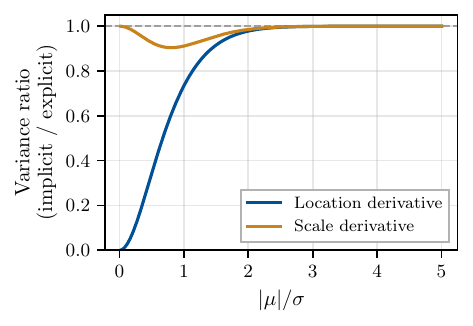}
  \caption{
    Variance ratio $\Var[\Gimp]/\Var[\Gexp]$ of the implicit $\FN$ estimator relative to the explicit estimator, as a function of the location-to-scale ratio $|\mu|/\sigma$, for the test function $f(Y) = Y^{2}$.
    Because the explicit variance is the upper bound, the ratio lies in $[0,1]$ and is independent of the number of Monte Carlo samples.
    The location-derivative ratio falls to $0$ at small $|\mu|/\sigma$, while the scale-derivative ratio dips only to $0.90$ near $|\mu|/\sigma \approx 0.8$.
  }
  \label{fig:fn_fractional_reduction}
\end{figure}

The explicit and implicit $\FN$ gradient estimators are unbiased, and their variances scale as $1/N$ with the number of Monte Carlo samples (Fig.~\ref{fig:th_vs_exp_varn}).
For the derivative with respect to $\mu$, the implicit estimator exhibits lower variance than the explicit estimator, and by Theorem~\ref{thm:variance_ordering} its variance is given by equation~\ref{eq:theo_var_fn_implicit} (Fig.~\ref{fig:th_vs_exp_varn}, left).
For the derivative with respect to $\sigma$, the empirical variances of the explicit estimator $\Gexpsigma$ and the implicit estimator $\Gimpsigma$ nearly overlap and match the theoretical prediction from equation~\ref{eq:fn_var_gexp} (Fig.~\ref{fig:th_vs_exp_varn}, right).
The variances of the explicit and implicit $\FN$ location-derivative estimators across a $(\mu, \sigma)$ parameter grid are shown in Fig.~\ref{fig:fn_imp_exp_dfdmu_var_subplots}, and the scale-derivative variances in Fig.~S1.
Figure~\ref{fig:fn_fractional_reduction} shows the variance ratio $\Var[\Gimp]/\Var[\Gexp]$, which is independent of the number of Monte Carlo samples $N$, and lies in $[0,1]$ since the explicit variance is an upper bound on the implicit variance.
The variance ratio of the location-derivative falls to $0$ as $|\mu|/\sigma \to 0$, while the scale-derivative ratio dips to $0.90$ near $|\mu|/\sigma \approx 0.8$.
We note that as $|\mu|/\sigma \to \infty$ both ratios approach $1$, implying that the estimators exhibit approximately the same variance.

\subsection{Variance of the Rice estimators}

\begin{figure}[h!tb]
  \centering
  \includegraphics[width=\textwidth]{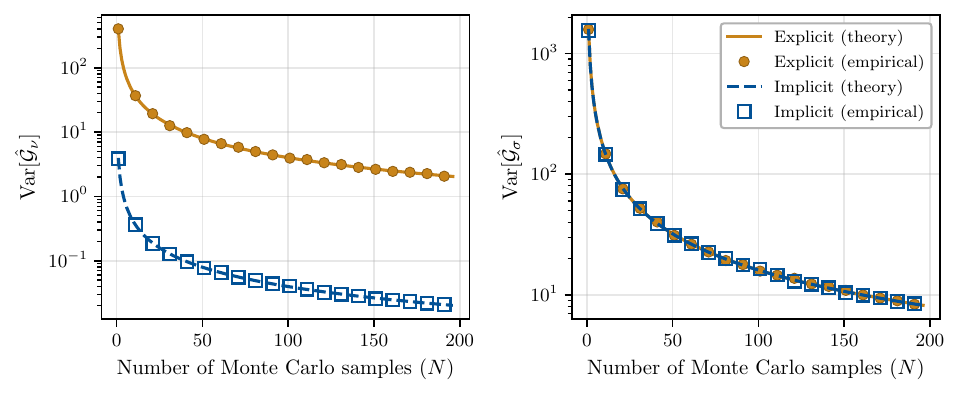}
  \caption{
    Theoretical and empirical variance of the $\Rice$ pathwise estimators as a function of Monte Carlo sample size $N$, at $\nu = 1, \sigma = 10$ (small $\nu/\sigma$).
    Left, the derivative with respect to $\nu$; right, the derivative with respect to $\sigma$.
    The two empirical curves match their respective theoretical predictions: $\Var[\Gexpnu] = 4\sigma^{2}/N$ for the explicit estimator and equation~\ref{eq:theo_var_rice_implicit} for the implicit estimator (left), where the expected conditional variance is the analytical gap from Theorem~\ref{thm:rice_variance_ordering}; for the $\sigma$ derivative (right) the theoretical curves coincide with the empirical curves.
    For the $\nu$ derivative at this parameter setting, the implicit estimator achieves variance roughly 100-fold smaller than the explicit estimator.
  }
  \label{fig:rice_th_vs_exp_varn}
\end{figure}

\begin{figure}[h!tb]
  \centering
  \includegraphics[width=\textwidth]{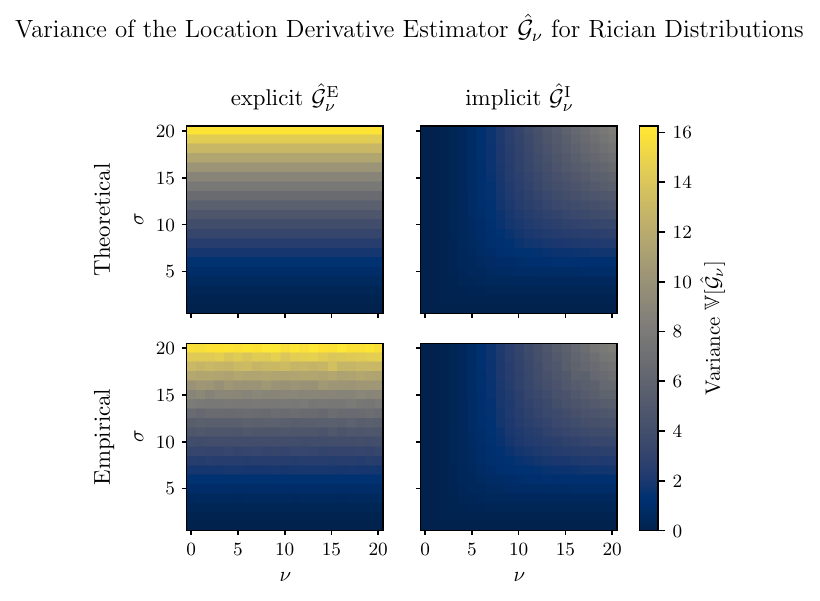}
  \caption{
    Theoretical (top) and empirical (bottom) variance of the explicit ($\Gexpnu$) and implicit ($\Gimpnu$) $\Rice$ derivative estimators of $\partial_{\nu}\E[R^{2}]$ as a function of $\nu$ and $\sigma$.
    The colorbar denotes the variance value for each $(\nu,\sigma)$ pair.
    The empirical variance was computed across $10{,}000$ simulations using $100$ Monte Carlo samples each.
    The theoretical variance was calculated using equations~\ref{eq:rice_var_gexp} and \ref{eq:theo_var_rice_implicit}, for the explicit and implicit estimators, respectively.
  }
  \label{fig:rice_imp_exp_dfdmu_var_subplots}
\end{figure}

\begin{figure}[h!tb]
  \centering
  \includegraphics{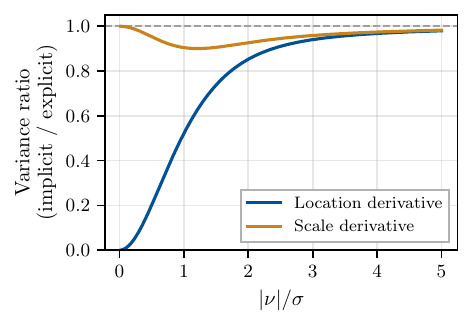}
  \caption{
    Variance ratio $\Var[\Gimp]/\Var[\Gexp]$ of the implicit $\Rice$ estimator relative to the explicit estimator, as a function of the location-to-scale ratio $\nu/\sigma$, for the test function $f(R) = R^{2}$.
    As in the $\FN$ case, the ratio lies in $[0,1]$ and is independent of the number of Monte Carlo samples.
    The location-derivative ratio falls to $0$ at small $\nu/\sigma$, while the scale-derivative ratio dips only to $0.90$ near $\nu/\sigma \approx 1.2$.
  }
  \label{fig:rice_fractional_reduction}
\end{figure}

The $\Rice$ estimators exhibit qualitatively the same pattern as the $\FN$ case, with both variances scaling as $1/N$ (Fig.~\ref{fig:rice_th_vs_exp_varn}).
For the derivative with respect to $\nu$, the implicit estimator exhibits lower variance than the explicit estimator, and by Theorem~\ref{thm:rice_variance_ordering} its variance is given by equation~\ref{eq:theo_var_rice_implicit}, with the largest improvement at small $\nu/\sigma$ (Fig.~\ref{fig:rice_th_vs_exp_varn}, left).
For the derivative with respect to $\sigma$, the explicit and implicit empirical variances nearly overlap and match the theoretical prediction (Fig.~\ref{fig:rice_th_vs_exp_varn}, right).
In both panels the empirical variances match their respective theoretical curves.
The variances of the explicit and implicit $\Rice$ location-derivative estimators across the $(\nu, \sigma)$ parameter grid are shown in Fig.~\ref{fig:rice_imp_exp_dfdmu_var_subplots}, and the scale-derivative variances in Fig.~S2.
As with the $\FN$ case, Fig.~\ref{fig:rice_fractional_reduction} summarizes the magnitude of the improvement as the variance ratio, which shows that the location-derivative ratio falls to $0$ as $\nu/\sigma \to 0$, while the scale-derivative ratio dips only to $0.90$ near $\nu/\sigma \approx 1.2$.

\subsection{French--Wilson scaling}
\label{subsec:fw_results}

We evaluate the variational French--Wilson model on a synthetic dataset of $2{,}240$ $(I_{\text{obs}}, \sigma_I, \Sigma)$ samples.
To obtain reference values, we evaluate the exact posterior by quadrature and take its mean and standard deviation as the reference amplitude $F$ and uncertainty $\sigma_F$.
We ask whether the choice of estimator, explicit or implicit, affects (1) how well each recovers the reference moments, and (2) the rate of convergence.
The model is trained using five different optimization algorithms: stochastic gradient descent \cite{1951robbins} with and without momentum \cite{1964polyak}, RMSprop \cite{2012tieleman}, Adam \cite{2015kingma} and AdaBelief \cite{2020zhuang}.
The model is also applied to an experimental dataset (PDB: 9B7C).

\subsubsection{Agreement with synthetic reference}
\label{subsubsec:fw_agreement}

We use the mean and standard deviation of the approximate posterior as point estimates for the amplitude and its uncertainty, which we denote by $\hat{F}$ and $\sigma_{\hat{F}}$, respectively.
Agreement between the point estimates and the reference is reported as the mean absolute percentage error ($100 \cdot |\hat{F} - F| / F$ and  $100 \cdot |\sigma_{\hat{F}} - \sigma_{F}| / \sigma_{F}$).

\begin{figure}[h!tb]
  \centering
  \includegraphics[width=\textwidth]{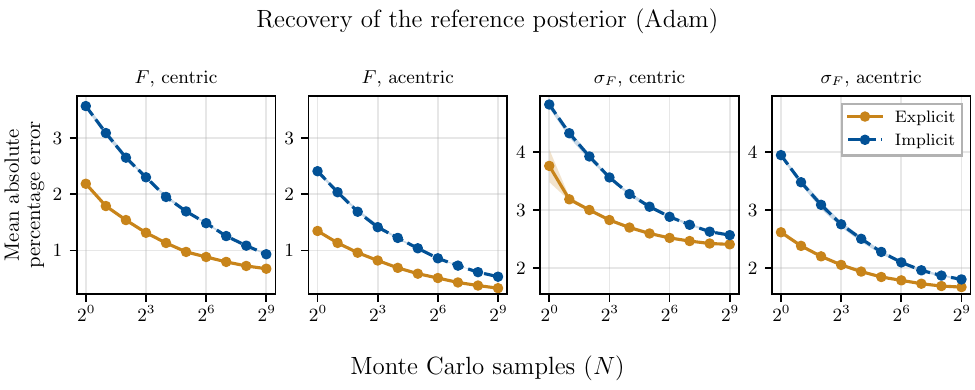}
  \caption{
    Recovery of the reference posterior, optimized with Adam, as a function of the number of Monte Carlo samples $N$ used per gradient.
    Panels show the amplitude $F$ and the posterior width $\sigma_F$, plotted separately for centric and acentric reflections.
    Each point is a median over ten replicates, and the bands span the interquartile range (too narrow to visualize).
    The same comparison for all five optimization algorithms is given in Fig.~S3.
  }
  \label{fig:fw_recovery}
\end{figure}

The variational approximation recovers both the mean and standard deviation of the reference posterior (Fig.~\ref{fig:fw_recovery}).
With the Adam optimizer and $N = 512$ Monte Carlo samples, the explicit estimator matches the reference centric amplitudes to $0.67\%$ and the acentric amplitudes to $0.33\%$, and the centric and acentric posterior standard deviations to $2.41\%$ and $1.67\%$.
The implicit estimator matches the same four quantities, in the same order, to $0.93\%$, $0.53\%$, $2.57\%$ and $1.81\%$.
All four errors decrease with an increasing number of Monte Carlo samples, and reach the lowest error at the largest $N$ we tested.
Table~S2 gives the same four quantities at $N = 512$ for every optimization algorithm, and Fig.~S3 shows their dependence on $N$.

Acentric amplitudes are recovered more accurately than centric ones at every $N$, under every optimization algorithm (Fig.~S3).
With respect to acentric reflections, the two estimators are indistinguishable under RMSprop and both forms of stochastic gradient descent; the implicit estimator is less accurate under Adam and AdaBelief.
For centric reflections, the implicit estimator is less accurate under RMSprop, Adam and AdaBelief; and the two estimators are indistinguishable under both forms of stochastic gradient descent.
Under Adam and AdaBelief, the difference in accuracy between the two estimators falls steadily as the number of Monte Carlo samples grows.

\subsubsection{Effect of the reparameterizations on training}
\label{subsubsec:fw_estimator_comparison}

We measured the ratio of the implicit to explicit gradient variance at the initial and final step of training, for both reparameterizations and all five optimizers.
Both estimators are evaluated at the same variational parameters.
The measurement was replicated ten times for each estimator/optimizer pairing.
We then binned the per-reflection ratios by the reflection's observed signal-to-noise ratio ($\snr$), and report the median and interquartile range in each bin (Figs.~S4 and~S5).

At initialization the centric location ratio ranges from $0.23$ to approximately $1.0$, over bins with $\snr \le 1$; and the acentric location ratio ranges from $0.61$ to $1.0$ over the same bins.
The variance ratio for the scale derivatives is approximately $1.0$ at the start and end of training.
By convergence, the centric location ratio rises from its initial ratio in bins with $\snr \le 1$.
This indicates that the parameters have moved to a regime where the implicit and explicit variances are similar.
Thus, the reduced variance of the implicit estimator is largest at the start of training for weak centric reflections, and that reduction disappears by the end of training.
We next ask whether the implicit estimator's lower variance at the start of training has any effect on the rate of convergence or on the agreement of the final estimates.

The two estimators reduce the error at the same rate over most of training (Fig.~S6).
Under plain stochastic gradient descent and stochastic gradient descent with momentum, the two curves coincide, differing by no more than $0.02$ percentage points at any recorded step.
For RMSprop, the two curves are indistinguishable early in training, differing by at most $0.03$ percentage points through step $1{,}000$.
Adam and AdaBelief use different learning rates because we tuned the learning rates for each optimizer before carrying out our experiments (Section~S5 of the supporting information).
The implicit estimator uses a learning rate $3.3$ times larger than the explicit estimator (Table~S1).
Although the implicit estimator uses a higher learning rate, the two estimators converge to nearly the same ELBO value, and the explicit estimator achieves a lower mean absolute percentage error.

\begin{figure}[h!tb]
  \centering
  \includegraphics[width=\textwidth]{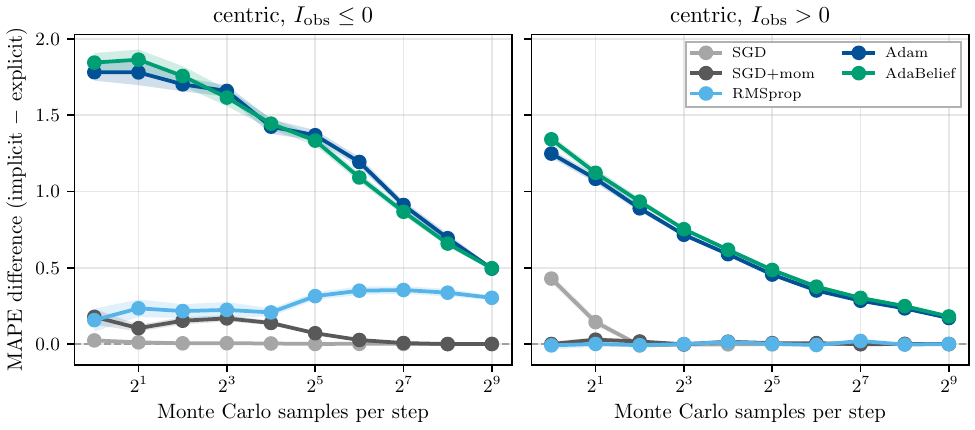}
  \caption{
    Difference in amplitude mean absolute percentage error (MAPE) against the reference posterior (implicit minus explicit), as a function of number of Monte Carlo samples used at each training step.
    Left, the centric reflections at or below zero intensity, where the two gradient variances differ; right, intensities with $\Iobs > 0$, where the implicit and explicit gradient variances are approximately the same.
    Positive values favor the explicit estimator at this number of optimization steps.
    Points are means over the ten per-replicate paired differences, and the bands are one standard error of the mean.
  }
  \label{fig:fw_gap_regime}
\end{figure}

The variance difference between the two estimators is largest for centric reflections with non-positive intensity ($I_\text{obs} \le 0$).
Therefore, we split the centric reflections into positive and non-positive intensities, and we plot the difference in amplitude recovery between the two estimators (Fig.~\ref{fig:fw_gap_regime}).
For both positive and non-positive intensities, we see that the two estimators agree more with each other as a function of increasing Monte Carlo samples $N$.
In the non-positive bin, where the variances differ the most, the explicit estimator is the more accurate of the two under the adaptive optimizers.
Under stochastic gradient descent, with and without momentum, the accuracy is nearly identical.
We also observe that the explicit estimator is more accurate in the positive bin under Adam and AdaBelief, while RMSprop and the stochastic gradient descent variants agree almost exactly.

The difference in accuracy between the explicit and implicit estimators (under the Adam optimizer) narrows with longer training, but it does not vanish.
We trained our model with the Adam optimizer for $200{,}000$ training steps, using only the centric reflections, and $N=512$ Monte Carlo samples.
The median difference in centric amplitude error is $0.47$ percentage points at $10{,}000$ steps, $0.27$ at $25{,}000$, and $0.25$ at $200{,}000$.
Most of the narrowing happens within the first $25{,}000$ steps, and the remaining $175{,}000$ close the gap by $0.03$ percentage points.

\subsubsection{Application to an experimental dataset}
\label{subsubsec:fw_realdata}

We used our model on a hen egg-white lysozyme dataset containing $45{,}196$ unique reflections that span a resolution range between $56.1$ and $1.10$\,\AA{} (PDB: 9B7C).
The dataset contains $4{,}791$ centric reflections, and $2.9\%$ of the reflections are non-positive ($I_{\text{obs}} \le 0$) (Section~S6 of the supporting information).
We binned the data into resolution bins to estimate the mean intensity $\Sigma$ of each bin for the Wilson prior (Section~S6 of the supporting information).
The fitted amplitudes are scored against numerical quadrature of the same posterior, as in Section~\ref{subsubsec:fw_agreement}.
Only the Adam optimizer was used because of its prevalence in modern machine learning algorithms.
A total of $25{,}000$ steps were used for training, and $N=512$ Monte Carlo samples were used for the gradient estimators.
As a sanity check, we also used the \prog{cctbx} French--Wilson scaling algorithm \cite{2002grossekunstleve} on the same set of reflections.

The variational posterior amplitudes agree with the quadrature to $0.12\%$ for the explicit estimator, and to $0.17\%$ for the implicit estimator (Table~\ref{tab:fw_realdata}).
The explicit estimator is more accurate (in terms of mean absolute percentage error) than the implicit, on acentric and centric reflections, for both the amplitude and uncertainty.
The \prog{cctbx} French--Wilson implementation agrees with the quadrature to $0.10\%$, and recovers the posterior standard deviations more closely than variational inference.

\begin{table}[h!tb]
  \centering
  \caption{
    The mean absolute percentage error of each method against the exact posterior, evaluated on an experimental hen egg-white lysozyme dataset (PDB: 9B7C).
  }
  \label{tab:fw_realdata}
  \begin{tabular}{lcccc}
    \toprule
    & \multicolumn{2}{c}{$F$ error (\%)} & \multicolumn{2}{c}{$\sigma_F$ error (\%)} \\
    \cmidrule(lr){2-3}\cmidrule(lr){4-5}
    Method          & centric & acentric & centric & acentric \\
    \midrule
    Explicit        & 0.22 & 0.11 & 1.76 & 1.54 \\
    Implicit        & 0.28 & 0.15 & 2.06 & 1.82 \\
    \prog{cctbx}  & 0.06 & 0.11 & 0.07 & 0.39 \\
    \bottomrule
  \end{tabular}
\end{table}

\section{Conclusion}

We implemented implicit and explicit gradient estimators of the $\FN$ and Rice distributions in \prog{PyTorch}, available in the open-source package \prog{rs-distributions}.
We validated the $\FN$ and $\Rice$ derivative estimators against analytical ground truths for the test functions $\E[Y^2]$ and $\E[R^2]$, respectively, and confirmed that the estimators are unbiased.
For both the $\FN$ and the Rice distributions, we proved that the implicit estimator is the conditional expectation of the explicit estimator given the sampled magnitude (Theorems~\ref{thm:variance_ordering} and~\ref{thm:rice_variance_ordering}).
Consequently, the law of total variance yields a variance bound that holds for any continuously differentiable function with finite-variance explicit gradient estimators.
The variance reduction is largest for the location derivative at small location-to-scale ratios, where the implicit estimator's variance drops by orders of magnitude relative to the explicit, while the scale-derivative variance falls by at most $10\%$ (Figs.~\ref{fig:fn_fractional_reduction} and~\ref{fig:rice_fractional_reduction}).
Our empirical variance measurements match our theoretical predictions for both distributions.

We then applied both estimators to French--Wilson scaling, where the $\FN$ and $\Rice$ distributions are the variational families for centric and acentric reflections.
The explicit estimator recovers a quadrature reference posterior to $0.67\%$ on the centric amplitude and $0.33\%$ on the acentric amplitude at $N = 512$ Monte Carlo samples per gradient.
The variance reduction is largest for centric reflections at or below zero intensity, which is in agreement with Theorems~\ref{thm:variance_ordering} and~\ref{thm:rice_variance_ordering}.
It is weaker on acentric reflections and near zero on the scale derivative of both classes.

We used our French--Wilson model to explore whether the reduced variance offered by the implicit estimator provided any practical advantages, such as a reduced error or faster convergence.
For this particular problem, the answer depends on the type of optimization algorithm used during training.
Under the adaptive optimizers, the explicit estimator is more accurate than the implicit.
However, this difference is reduced as a function of increasing number of Monte Carlo samples used to estimate the gradients.
Under stochastic gradient descent, neither of the estimators exhibits an advantage over the other.
On an experimental lysozyme dataset, both estimators reproduce the exact posterior amplitude to better than $1\%$.

\clearpage
\appendix
\section*{\centering Appendices}
\section{Proof of Theorem~\ref{thm:variance_ordering}}
\label{appendix:var_proof}

The key observation in this proof is that the implicit $\FN$ pathwise derivative is the Rao--Blackwellization of the explicit derivative with respect to the sampled magnitude \(Y = |X|\); once this identity is established, unbiasedness and the variance ordering follow immediately from standard conditional-expectation identities. Throughout this section, we let $a = (y - \mu)/\sigma$, $b = (y + \mu)/\sigma$, and we let $\phi$ denote the standard $\N$ PDF.

\textbf{Conditional distribution of $X$ given $Y$.}
The absolute-value function $Y = |X|$ is two-to-one for $Y > 0$, which means that the equation $|X| = y$ has solutions $X = +y$ and $X = -y$.
By Bayes' rule, with $p_{X}$ denoting the $\N(\mu,\sigma^{2})$ density,
\begin{equation}
  \mathbb{P}(X = +y \mid Y = y) = \frac{p_{X}(+y)}{p_{X}(+y) + p_{X}(-y)} = \frac{\phi(a)}{\phi(a) + \phi(b)} =: p_{+},
\end{equation}
and similarly $\mathbb{P}(X = -y \mid Y = y) = \phi(b) / [\phi(a) + \phi(b)] =: p_{-}$.

\textbf{Conditional expectation of the explicit gradient.}
Under explicit reparameterization, $\partial Y/\partial \mu = \sign(X)$ and $\partial Y/\partial \sigma = \sign(X) \cdot \epsilon = \sign(X) \cdot (X - \mu)/\sigma$.
Conditional on $Y = y$, each partial derivative takes on two values,
\begin{align}
  \partial_{\mu} Y \mid (X = +y) &= +1, & \partial_{\mu} Y \mid (X = -y) &= -1, \\
  \partial_{\sigma} Y \mid (X = +y) &= +a, & \partial_{\sigma} Y \mid (X = -y) &= +b.
\end{align}

Applying the Law of the Unconscious Statistician \cite{2020grimmett} to the expectation of the conditional distribution gives us
\begin{equation}
  \E[\partial_{\mu} Y \mid Y = y] = p_{+} \cdot (+1) + p_{-} \cdot (-1) = \frac{\phi(a) - \phi(b)}{\phi(a) + \phi(b)},
\end{equation}
and
\begin{equation}
  \E[\partial_{\sigma} Y \mid Y = y] = p_{+} \cdot a + p_{-} \cdot b = \frac{(y - \mu)\phi(a) + (y + \mu)\phi(b)}{\sigma[\phi(a) + \phi(b)]}.
\end{equation}
These match the implicit gradient formulas derived in Section~S2 of the supporting information for $f(Y) = Y^{2}$.
Since we required that $f$ is continuously differentiable, we know that the derivative $f'(Y)$ exists.
Further, $f'(Y)$ is fully determined conditional on $Y$, so it factors out of the conditional expectation,
\begin{equation}
  \label{eq:fn_rb_identity_proof}
  \E[\Gexptheta \mid Y] = \E[f'(Y) \cdot \partial_{\theta} Y \mid Y] = f'(Y) \cdot \E[\partial_{\theta} Y \mid Y] = \Gimptheta,
\end{equation}
where the final equality uses the conditional expectations of $\partial_{\theta} Y$ computed above, which shows the identity (\ref{eq:rb_identity}) stated in the main text.

\textbf{Unbiasedness and variance ordering.}

Both arguments below require the explicit estimator to have finite second moment, which is satisfied with the assumption $\E[(f'(Y))^{2}(1 + Y^{2})] < \infty$.
The unbiasedness of $\Gimptheta$ follows from the tower property \cite{2020grimmett} applied to (\ref{eq:fn_rb_identity_proof}),
\begin{equation}
  \E[\Gimptheta] = \E\big[\E[\Gexptheta \mid Y]\big] = \E[\Gexptheta] = \partial_{\theta}\E[f(Y)],
\end{equation}
where the last equality follows from the unbiasedness of the explicit estimator (Section~S1 of the supporting information).
The variance ordering (\ref{eq:variance_ordering}) follows from the law of total variance,
\begin{equation}
  \Var[\Gexptheta] = \Var\big[\E[\Gexptheta \mid Y]\big] + \E\big[\Var[\Gexptheta \mid Y]\big],
\end{equation}
and from the non-negativity of the second term.
The first term equals $\Var[\Gimptheta]$ by (\ref{eq:fn_rb_identity_proof}).
Equality holds if and only if $\Var[\Gexptheta \mid Y] = 0$.
$\square$

\section{Variance gap of the FoldedNormal estimators}
\label{appendix:fn_gap}

Applying the law of total variance to $\Gexptheta$ conditioned on $Y$, and using equation~\ref{eq:fn_rb_identity_proof}, expresses the gap between the two variances as an expected conditional variance,
\begin{align}
  \label{eq:gap_lotv}
  \Var[\Gexptheta] - \Var[\Gimptheta]
  \;=\; \E\big[\Var[\Gexptheta \mid Y]\big].
\end{align}
For the test function $f(Y) = Y^2$ and parameter $\theta = \{\mu, \sigma\}$, the conditional variance $\Var[\Gexptheta \mid Y]$ evaluates to
\begin{align}
  \Var[\Gexpmu \mid Y] &= \frac{16 Y^{2}\phi(a)\phi(b)}{[\phi(a) + \phi(b)]^{2}}, &
  \Var[\Gexpsigma \mid Y] &= \frac{16 Y^{2}\mu^{2}\phi(a)\phi(b)}{\sigma^{2}[\phi(a) + \phi(b)]^{2}}.
\end{align}
Taking the expectation with respect to the $\FN$ density $p_{Y}(y) = [\phi(a) + \phi(b)]/\sigma$ (equation~\ref{eq:fn_pdf}) yields the per-sample variance gaps
\begin{align}
  \label{eq:gap_mu}
  \E\big[\Var[\Gexpmu \mid Y]\big] &= \int_{0}^{\infty} \frac{16 y^{2}\phi(a)\phi(b)}{\sigma[\phi(a) + \phi(b)]}dy, \\
  \label{eq:gap_sigma}
  \E\big[\Var[\Gexpsigma \mid Y]\big] &= \int_{0}^{\infty} \frac{16\mu^{2}y^{2}\phi(a)\phi(b)}{\sigma^{3}[\phi(a) + \phi(b)]}dy.
\end{align}
These integrals have no closed form, but they can be evaluated by numerical quadrature.
Combining equation~\ref{eq:gap_lotv} with the closed-form explicit variances (equation~\ref{eq:fn_var_gexp}) yields the variance of the implicit estimator for $N$ Monte Carlo samples, equation~\ref{eq:theo_var_fn_implicit}.

\section{Proof of Theorem~\ref{thm:rice_variance_ordering}}
\label{appendix:rice_var_proof}

The argument mirrors the steps used in the proof of Theorem~\ref{thm:variance_ordering} (Appendix~\ref{appendix:var_proof}). We first establish the conditional density of the direction given $R$, then show that the conditional expectation of the explicit derivative recovers the implicit derivative, and finally apply the tower property and the law of total variance.
Write $\hat{\vz} = \vz/R$ and $\hat{\vmu} = \vmu/\nu$ for the unit sample and mean directions, respectively, and parametrize $\vz = R(\cos\Phi, \sin\Phi)$ with the polar axis along $\hat{\vmu}$, so that $\Phi$ is the angle between $\hat{\vz}$ and $\hat{\vmu}$ and $\hat{\vz}\cdot\hat{\vmu} = \cos\Phi$.

\textbf{Conditional distribution of the direction given $R$.}
The bivariate distribution $\vz \sim \N(\vmu, \sigma^2\vI)$ that defines $R = \norm{\vz}$ has density $p(\vz) \propto \exp\!\left(-\dfrac{\norm{\vz - \vmu}^2}{2\sigma^2}\right)$.
Restricting the density to a circle of radius $\norm{\vz} = R$, and using $\norm{\vz - \vmu}^{2} = R^{2} + \nu^{2} - 2R\nu\cos\Phi$, the $R^{2} + \nu^{2}$ terms are constant in $\Phi$, so $p(\Phi \mid R) \propto \exp\big(\tfrac{R\nu}{\sigma^{2}}\cos\Phi\big) = \exp(\kappa\cos\Phi)$.
Normalizing over $\Phi \in [0, 2\pi]$ gives a von Mises density \cite{1997bricogne},
\begin{equation}
  \label{eq:rice_angle_density}
  p(\Phi \mid R) = \frac{\exp(\kappa\cos\Phi)}{2\pi I_{0}(\kappa)},
\end{equation}
with concentration $\kappa = \frac{\nu R}{\sigma^{2}}$, where $I_{0}$ is the modified Bessel function of the first kind \cite{2009mardia}.
This is the $\Rice$ analogue of the two-point Bayes step in Appendix~\ref{appendix:var_proof}.
In that case, the pre-image values, $\{+y, -y\}$, were weighted by $\phi(a)$ and $\phi(b)$.
Here, the pre-image values lie on the circle of radius $R$, and they are weighted by the von Mises density.

We note that the mean direction $\hat \vmu = \vmu / \nu$ and the angle $\Phi$ require that $\nu > 0$.
When $\nu = 0$, the mean vanishes, $\vz \sim \N(\vzero, \sigma^2\vI)$ is isotropic, and $R$ is Rayleigh.
Further, when $\nu = 0$ we get that $\kappa = 0$, so that $I_0(0) = 1$, and the von Mises density is uniform $p(\Phi \mid R) = 1/(2\pi)$.

\textbf{Conditional expectation of the explicit gradient.}
The pathwise derivatives of $R$ were obtained in vector form in Section~S3 of the supporting information, and are given by
\begin{align}
  \partial_{\nu} R = \hat{\vz}\cdot\hat{\vmu} = \cos\Phi, \qquad
  \partial_{\sigma} R = \hat{\vz}\cdot\veps = \frac{R - \nu\cos\Phi}{\sigma}.
\end{align}
Both terms are affine, and conditioning on $R$ fixes the magnitude, leaving $\Phi$ as the only random variable.
Hence, by linearity of expectation, the only average required for each term is $\E[\cos\Phi \mid R]$.
Applying the Law of the Unconscious Statistician \cite{2020grimmett} with the von Mises density,
\begin{equation}
  \E[\cos\Phi \mid R] = \int_{0}^{2\pi}\cos\Phi\;p(\Phi \mid R)d\Phi
  = \frac{\int_{0}^{2\pi}\cos\Phi e^{\kappa\cos\Phi} d\Phi}{2\pi I_{0}(\kappa)}.
\end{equation}
Because the von Mises density is an exponential family with natural parameter $\kappa$ and sufficient statistic $\cos\Phi$, this mean equals the derivative of the log-normalizer $\log\big(2\pi I_{0}(\kappa)\big)$ \cite{2008wainwright},
\begin{equation}
  \E[\cos\Phi \mid R] = \frac{d}{d\kappa}\log\big(2\pi I_{0}(\kappa)\big) = \frac{I_{0}'(\kappa)}{I_{0}(\kappa)} = \frac{I_{1}(\kappa)}{I_{0}(\kappa)},
\end{equation}
where the final step uses $I_{0}' = I_{1}$.
Substituting the conditional expectation back into the derivatives gives us
\begin{align}
  \E[\partial_{\nu} R \mid R] = \frac{I_{1}(\kappa)}{I_{0}(\kappa)}, \qquad
  \E[\partial_{\sigma} R \mid R] = \frac{1}{\sigma}\left(R - \nu \frac{I_{1}(\kappa)}{I_{0}(\kappa)}\right).
\end{align}
These are exactly the implicit derivative terms $\partial_{\nu} R$ and $\partial_{\sigma} R$ obtained from the Marcum-$Q$ CDF in Section~S4 of the supporting information, so the conditional-expectation route and the CDF route produce the same $I_{1}/I_{0}$ term.
Because we required $f$ to be continuously differentiable, the derivative $f'(R)$ exists, and as a function of $R$ alone it factors out of the conditional expectation,
\begin{equation}
  \label{eq:rice_rb_identity_proof}
  \E[\Gexptheta \mid R] = f'(R) \E[\partial_{\theta} R \mid R] = \Gimptheta,
\end{equation}
which establishes the identity (\ref{eq:rice_rb_identity}) stated in the main text.

\textbf{Unbiasedness and variance ordering.}
Both arguments below require the explicit estimator to have finite second moment, which the requirement $\E[(f'(R))^{2}(1 + R^{2})] < \infty$ provides.
The unbiasedness of $\Gimptheta$ follows from the tower property \cite{2020grimmett} applied to (\ref{eq:rice_rb_identity_proof}),
\begin{equation}
  \E[\Gimptheta] = \E\big[\E[\Gexptheta \mid R]\big] = \E[\Gexptheta] = \partial_{\theta} \E[f(R)],
\end{equation}
where the last equality uses the unbiasedness of the explicit estimator (Section~S3 of the supporting information).
The variance ordering (\ref{eq:rice_variance_ordering}) follows from the law of total variance,
\begin{equation}
  \Var[\Gexptheta] = \Var\big[\E[\Gexptheta \mid R]\big] + \E\big[\Var[\Gexptheta \mid R]\big],
\end{equation}
and the non-negativity of the second term.
The first term equals $\Var[\Gimptheta]$ by (\ref{eq:rice_rb_identity_proof}).
Equality holds if and only if $\Var[\Gexptheta \mid R] = 0$.
$\square$

\section{Variance gap of the Rice estimators}
\label{appendix:rice_gap}

Conditional on $R$, the direction $\Phi$ follows the von Mises density (Appendix~\ref{appendix:rice_var_proof}), an exponential family with natural parameter $\kappa = \nu R/\sigma^{2}$ and sufficient statistic $\cos\Phi$.
The variance of the sufficient statistic is therefore the second derivative of the log-normalizer, $\Var[\cos\Phi \mid R] = \frac{d^{2}}{d\kappa^{2}}\log I_{0}(\kappa)$ \cite{2008wainwright}.
Differentiating $\frac{d}{d\kappa}\log I_{0}(\kappa) = I_{1}(\kappa)/I_{0}(\kappa)$ once more with the quotient rule gives
\begin{equation}
  \label{eq:rice_cond_var_deriv}
  \Var[\cos\Phi \mid R] = \frac{d}{d\kappa}\frac{I_{1}(\kappa)}{I_{0}(\kappa)} = \frac{I_{1}'(\kappa)}{I_{0}(\kappa)} - \left(\frac{I_{1}(\kappa)}{I_{0}(\kappa)}\right)^{2}.
\end{equation}
Using $I_{0}' = I_{1}$ and $I_{1}' = \tfrac{1}{2}(I_{0} + I_{2})$ in the first term gives
\begin{equation}
  \label{eq:rice_cond_var_nu_proof}
  \Var[\cos\Phi \mid R] = \frac{1}{2}\left(1 + \frac{I_{2}(\kappa)}{I_{0}(\kappa)}\right) - \left(\frac{I_{1}(\kappa)}{I_{0}(\kappa)}\right)^{2}.
\end{equation}

Now, applying the law of total variance to $\Gexptheta$ conditioned on $R$, and using equation~\ref{eq:rice_rb_identity_proof}, expresses the gap between the two variances as an expected conditional variance,
\begin{equation}
  \label{eq:rice_gap_lotv}
  \Var[\Gexptheta] - \Var[\Gimptheta] = \E\big[\Var[\Gexptheta \mid R]\big].
\end{equation}
For the test function $f(R)= R^2$ and the parameter $\theta = \{ \nu, \sigma\}$, the conditional variance $\Var[\Gexptheta \mid R]$ is given by the terms
\begin{align}
  \Var[\Gexpnu \mid R] = 4R^{2}\Var[\cos\Phi \mid R], \quad \text{and} \quad
  \Var[\Gexpsigma \mid R] = \frac{4\nu^{2}R^{2}}{\sigma^{2}}\Var[\cos\Phi \mid R],
\end{align}
where $\Var[\cos\Phi \mid R]$ is given by equation \ref{eq:rice_cond_var_nu_proof}.
Taking the expectation of $\Var[\Gexpnu \mid R]$ with respect to the $\Rice$ density $p(r;\nu,\sigma)$ (equation~\ref{eq:rice_pdf}) yields the per-sample variance gaps
\begin{equation}
  \label{eq:rice_gap_nu}
  \E[\Var[\Gexpnu \mid R]] = \int_{0}^{\infty} 4r^{2}\Var[\cos\Phi \mid R{=}r]p(r;\nu,\sigma)dr,
\end{equation}
and
\begin{equation}
  \label{eq:rice_gap_sigma}
  \E[\Var[\Gexpsigma \mid R]] = \frac{\nu^{2}}{\sigma^{2}}\ \int_{0}^{\infty} 4r^{2}\Var[\cos\Phi \mid R{=}r]p(r;\nu,\sigma)dr.
\end{equation}
The remaining integral in the variance terms involves ratios of modified Bessel functions and has no closed form.
As in the $\FN$ case, we can evaluate the integral by numerical quadrature.
Combining equation~\ref{eq:rice_gap_lotv} with the closed-form explicit variances $\Var[\Gexpnu] = 4\sigma^{2}/N$ and $\Var[\Gexpsigma] = 4(\nu^{2} + 4\sigma^{2})/N$ (Section~\ref{sssec:rice_explicit_estimator}) yields the variance of the implicit estimator for $N$ Monte Carlo samples, equation~\ref{eq:theo_var_rice_implicit}.

\begin{acknowledgements}
  We would like to thank Doris Mai for contributing to and helping maintain our open-source package \textit{rs-distributions}.

  This work depends on open-source software, and we thank the developers and maintainers of \prog{PyTorch} \cite{2019paszke}, \prog{ArviZ} \cite{2019kumar}, \prog{cctbx} \cite{2002grossekunstleve}, \prog{reciprocalspaceship} \cite{2021greisman}, \prog{gemmi} \cite{2022wojdyr}, \prog{adabelief-pytorch}, \prog{NumPy} \cite{2020harris}, \prog{SciPy} \cite{2020virtanen}, \prog{pandas}, \prog{Polars}, \prog{Matplotlib} and \prog{seaborn}.

  We used Claude Code (Anthropic), running Claude Opus 5, as a coding assistant while writing the analysis and figure-generation scripts, and for editing the text.
  The code and text were checked for errors, and the authors take full responsibility for the content of this paper.
\end{acknowledgements}

\begin{funding}
  This work was supported by National Institutes of Health (NIH) grant DP2-GM141000 (to D.R.H.), National Science Foundation Graduate Research Fellowship grant 2023321426 (to L.A.A.).
  The Linac Coherent Light Source (LCLS) at the SLAC National Accelerator Laboratory is an Office of Science User Facility operated for the U.S. Department of Energy Office of Science by Stanford University. This work was supported by the U.S. Department of Energy, Office of Science, Basic Energy Sciences under Contract No. DEAC02-76SF00515.
\end{funding}

\ConflictsOfInterest{The authors declare there are no conflicts of interest.
}

\DataAvailability{
  The implicit and explicit $\FN$ and $\Rice$ distributions are implemented in the open-source \prog{PyTorch} package \prog{rs-distributions}, at \url{https://github.com/rs-station/distributions}.
  The code that reproduces every analysis, figure and table in this work, including the French--Wilson variational model from Section~S5 of the supporting information, is at \url{https://github.com/Hekstra-Lab/gradient_estimator_paper}.
  That repository also contains the script that generates the synthetic dataset from Section~S6 of the supporting information.
  The experimental data are the deposited structure factors of PDB entry 9B7C, available from the Protein Data Bank at \url{https://www.rcsb.org/structure/9B7C}.
}

\bibliography{iucr}

\clearpage
\setcounter{section}{0}
\setcounter{figure}{0}
\setcounter{table}{0}
\setcounter{equation}{0}
\setcounter{algorithm}{0}
\renewcommand{\thesection}{S\arabic{section}}
\renewcommand{\thefigure}{S\arabic{figure}}
\renewcommand{\thetable}{S\arabic{table}}
\renewcommand{\theequation}{S\arabic{equation}}
\renewcommand{\thealgorithm}{S\arabic{algorithm}}
\renewcommand{\theHsection}{S\arabic{section}}
\renewcommand{\theHfigure}{S\arabic{figure}}
\renewcommand{\theHtable}{S\arabic{table}}
\renewcommand{\theHequation}{S\arabic{equation}}
\renewcommand{\theHalgorithm}{S\arabic{algorithm}}
\section*{\centering Supporting information}

\section{Derivation of the FoldedNormal explicit gradient estimator}
\label{appendix:fn_explicit}

Let $f(Y) = Y^2$ so that our loss function is $\E[f(Y)]$.
We are interested in the variance and expectation of the gradient estimator for
\begin{equation}
  \begin{aligned}
    \nabla_{\theta} \E_{\epsilon \sim \N(0,1)}[f(Y)] & = \E_{\epsilon \sim
    \N(0,1)} [ \nabla _\theta f(Y)]
    \\
    & \approx \frac{1}{N} \sum _{n=1}^N \nabla _\theta f(Y) \qquad
    (\text{Monte Carlo sampling})
  \end{aligned}
\end{equation}
For a single Monte Carlo sample, $n = 1$, we have
\begin{align}
  \nabla_{\theta} f(Y) & =  f'(Y) \frac{\partial Y}{\partial X} \nabla_\theta X    .
\end{align}
The first component is given by
\begin{align}
  f'(Y) = 2Y.
\end{align}
The second component is
\begin{equation}
  \begin{aligned}
    \frac{\partial Y} {\partial X} & = \frac{\partial | X |
    }{\partial X}
    \\
    & = \sign(X) \qquad \text{(for $X \neq 0$)} \\
  \end{aligned}
\end{equation}
The third component involves $X = \mu + \sigma \epsilon$, so we have to
differentiate with respect to $\mu$ and $\sigma$.
Differentiating with respect to $\mu$ gives
\begin{align}
  \frac{\partial X}{\partial \mu} & = 1,
\end{align}
and differentiating with respect to $\sigma$ gives us
\begin{align}
  \frac{\partial X}{\partial \sigma} & = \epsilon.
\end{align}

Hence, we get that
\begin{equation}
  \begin{aligned}
    \frac{\partial f(Y)}{\partial \mu} & = 2Y \cdot \sign(X) \cdot 1
    \\
    & = 2Y \cdot \sign(X)                                   \\
    & = 2 |X| \cdot   \sign (X)                             \\
    & =  2X \qquad \text{(since $|X| \cdot \sign(X) = X$)},
  \end{aligned}
\end{equation}
and
\begin{equation}
  \begin{aligned}
    \frac{\partial f(Y)}{\partial \sigma} & = 2Y \cdot \sign(X) \cdot \epsilon \\
    & = 2X\epsilon                       \\
    & = 2(\mu + \sigma \epsilon)\epsilon        \\
    & = 2\mu \epsilon + 2 \sigma \epsilon^2.
  \end{aligned}
\end{equation}
These are the explicit derivative estimators when using a single Monte Carlo sample, for the test function $f(Y) = Y^2$.

Now we find the variance and the expectation of the gradients:
\begin{equation}
  \begin{aligned}
    \E\left[\frac{\partial f(Y)}{\partial \mu}\right] & =\E[2X]
    \\
    & = 2 \E[X] \qquad \text{(by linearity of expectation)}          \\
    & = 2\mu \qquad \text{(since $X \sim \N(\mu,\sigma^2)$)},
  \end{aligned}
\end{equation}
and
\begin{equation}
  \begin{aligned}
    \label{eq:expectation_dfdsig}
    \E\left[\frac{\partial f(Y)}{\partial \sigma}\right] & =\E[2 \mu
    \epsilon + 2 \sigma \epsilon^2 ]
    \\
    & = 2 \mu \E[\epsilon] + 2 \sigma \E[\epsilon^2] \qquad \text{(by linearity of
    expectation)}
    \\
    & = 0 + 2\sigma \qquad \text{(since $\E[\epsilon^2] = \Var[\epsilon] +
    (\E[\epsilon])^2 = 1$)}
    \\
    & = 2 \sigma.
  \end{aligned}
\end{equation}
Hence, the estimators are unbiased.
The variance is given by:
\begin{equation}
  \begin{aligned}
    \Var \left[\frac{\partial f(Y)}{\partial \mu}\right] & = \Var[2X]    \\
    & = 4\Var[X]    \\
    & = 4 \sigma^2,
  \end{aligned}
\end{equation}
and
\begin{equation}
  \begin{aligned}
    \Var \left[\frac{\partial f(Y)}{\partial \sigma}\right] & =
    \Var[2X\epsilon]
    \\
    & = 4\Var[X\epsilon]                   \\
    & = 4(\E[X^2\epsilon^2] - (\E[X\epsilon])^2).
  \end{aligned}
\end{equation}
From equation~\ref{eq:expectation_dfdsig}, we know that $\E[X\epsilon] = \sigma $.
For the term $(\E[X^2\epsilon^2])$, we get
\begin{equation}
  \begin{aligned}
    \E[X^{2}\epsilon^{2}] & = \E[(\mu + \sigma \epsilon)^{2} \epsilon^{2}]
    \\
    & = \E[\mu^2\epsilon^{2} +2 \mu \sigma \epsilon^3 + \sigma^{2} \epsilon^{4}]          \\
    & =  \mu^2\E[\epsilon^2] + 2 \mu \sigma \E[\epsilon^{3}] + \sigma^{2}\E[\epsilon^{4}] \\
    & = \mu^{2}(1) + 2 \mu \sigma (0) + \sigma^{2} ( 3)              \\
    & = \mu^{2} + 3 \sigma^{2}.
  \end{aligned}
\end{equation}
Hence, we get
\begin{equation}
  \begin{aligned}
    \Var \left[\frac{\partial f(Y)}{\partial \sigma}\right] & =
    4(\mu^{2} + 3 \sigma^{2} - \sigma^{2})
    \\
    & = 4(\mu^{2} + 2\sigma^{2}).
  \end{aligned}
\end{equation}
From these derivations we see that the variance of the estimator with respect to $\mu$ depends on $\sigma$, and the variance of the estimator with respect to $\sigma$ depends on both $\mu$ and $\sigma$.
In practice, we use more than one Monte Carlo sample ($N > 1$), so the expectation becomes
\begin{equation}
  \begin{aligned}
    \E\left[\frac{1}{N}\sum_{n=1}^{N} \nabla _{\theta}f(Y)\right] & =
    \frac{1}{N} \sum_{n=1}^{N} \E\left[\nabla _{\theta}f(Y)\right]
    \\
    & = \E \left[\nabla _{\theta}f(Y)\right],
  \end{aligned}
\end{equation}
which is the same as the single-sample case because of the linearity of expectation.
If we assume that each Monte Carlo sample is i.i.d., we get the following variance
\begin{equation}
  \begin{aligned}
    \Var\left[\frac{1}{N} \sum_{n=1}^{N} \nabla_{\theta}f (Y)\right]
    & = \frac{1}{N^{2}}\sum_{n=1}^{N} \Var[\nabla _{\theta}f(Y)] \\
    & = \frac{1}{N} \Var [\nabla _{\theta}f(Y)].
  \end{aligned}
\end{equation}
When we differentiate with respect to $\mu$ we get
\begin{flalign}
  \frac{1}{N} \Var [\partial_{\mu} f(Y)] & = \frac{4 \sigma^{2}}{N}.
\end{flalign}
Differentiating with respect to $\sigma$ gives us
\begin{flalign}
  \frac{1}{N}\Var[\partial_{\sigma}f(Y)] = \frac{4(\mu^{2} + 2\sigma^{2})}{N}.
\end{flalign}

\section{Derivation of the FoldedNormal implicit gradient estimator}
\label{appendix:fn_implicit}

Starting from equation~5 of the main text, we use the CDF of the $\FN$ distribution as the standardization function $s$.
The CDF of the $\FN$, denoted here by $\cdf$, has the form
\begin{equation}
  \begin{aligned}
    F(Y;\theta) & = \frac{1}{2}\left[\erf\left(\frac{Y +
      \mu}{\sigma\sqrt{2}}\right) + \erf\left(\frac{Y -
    \mu}{\sigma\sqrt{2}}\right)\right],
  \end{aligned}
\end{equation}
where $\erf$ is the error function.
Differentiating $\cdf$ with respect to $\mu$ gives us
\begin{equation}
  \begin{aligned}
    \frac{\partial}{\partial \mu} \cdf & =  \frac{1}{\sqrt{2 \pi \sigma^{2}} }
    \exp\left(\frac{-(Y + \mu)^{2}}{2 \sigma^{2}}\right) -
    \frac{1}{\sqrt{2 \pi \sigma^{2}} } \exp\left(\frac{-(Y-
    \mu)^{2}}{2 \sigma^{2}}\right).
  \end{aligned}
  \label{eq:dCDF_du}
\end{equation}
We note that the two components on the right-hand side of equation~\ref{eq:dCDF_du} are Normal distribution PDFs evaluated at $Y$, with means $-\mu$ and $\mu$ respectively (both with scale $\sigma$).
Then, differentiating $\cdf$ with respect to $\sigma$ gives us
\begin{equation}
  \begin{aligned}
    \frac{\partial}{\partial \sigma}\cdf & = -\frac{(Y- \mu)}{\sigma}
    \frac{1}{\sqrt{2 \pi \sigma^{2}}} \exp \left( \frac{- (Y -
    \mu)^{2}}{2 \sigma ^{ 2}}\right) - \frac{(Y +
    \mu)}{\sigma}\frac{1}{\sqrt{2 \pi \sigma^{2}}}\exp \left(\frac{- (Y
    + \mu)^{2}}{2 \sigma^{2}}\right).
  \end{aligned}
  \label{eq:dCDF_dsig}
\end{equation}
Defining the following variables,
\begin{gather}
  a = \frac{Y - \mu}{\sigma}
  \quad\text{and}\quad
  b = \frac{Y + \mu}{\sigma},
\end{gather}
and letting $\phi$ denote the PDF of the standard Normal
distribution, we can write shorter expressions for equations~\ref{eq:dCDF_du} and \ref{eq:dCDF_dsig},
\begin{equation}
  \begin{aligned}
    \frac{\partial}{\partial \mu} \cdf & =\frac{1}{\sigma}\left( \phi(b) - \phi(a)\right)   \\
    & = -\frac{1}{\sigma}\left(\phi(a) - \phi(b)\right),
  \end{aligned}
\end{equation}
and
\begin{equation}
  \begin{aligned}
    \frac{\partial}{\partial \sigma} \cdf & = - \frac{(Y - \mu)}{\sigma^{2}}\phi(a) -
    \frac{Y + \mu}{\sigma^{2}} \phi(b)
    \\
    & = -\frac{1}{\sigma^2}\left((Y - \mu)\phi(a) + (Y + \mu)\phi(b)\right).
  \end{aligned}
\end{equation}
This allows us to write the per-sample implicit gradients as
\begin{equation}
  \begin{aligned}
    \partial_{\mu} f(Y) & = f'(Y) \left(- \frac{\partial_{\mu}
    F(Y;\theta)}{p(Y)}\right)                                             \\
    & = 2Y  \frac{\phi(a) - \phi(b)}{\phi(a) + \phi(b)},
  \end{aligned}
\end{equation}
and
\begin{equation}
  \begin{aligned}
    \partial_{\sigma} f(Y) & = f'(Y)\left(- \frac{\partial_{\sigma}
    F(Y;\theta)}{p(Y)}\right)                                              \\
    & = \frac{2Y}{\sigma}\frac{(Y - \mu)\phi(a) + (Y +
    \mu)\phi(b)}{\phi(a) + \phi(b)}.
  \end{aligned}
\end{equation}
Averaging these per-sample gradients over $N$ Monte Carlo samples yields the estimators $\Gimpmu$ and $\Gimpsigma$ reported in the main text.

\section{Derivation of the Rice explicit gradient estimator}
\label{appendix:rice_explicit}

Let $R = \norm{\vz} \sim \Rice(\nu,\sigma)$, where $\vz = \vmu + \sigma\veps$ with $\vmu = (\mu_{1},\mu_{2})$, $\nu = \lVert\vmu\rVert$, and $\epsilon_1,\epsilon_2 \overset{\text{i.i.d.}}{\sim} \N(0,1)$.
Using the explicit pathwise Monte Carlo estimator shown in equation~4 of the main text, we want to find the per-sample gradient $\nabla_{\theta}f(R)$, where $f(R) = R^2$.
Differentiating with respect to $\mu_{1}$, we get
\begin{equation}
  \begin{aligned}
    \partial_{\mu_{1}}f(R) & = \partial_{\mu_{1}} R^2
    \\
    & = \partial_{\mu_{1}} (z_{1}^{2} + z_{2}^{2})
    \\
    & = \partial_{\mu_{1}} (\mu_{1} + \sigma\epsilon_1)^{2} +
    \partial_{\mu_{1}}(\mu_{2} + \sigma\epsilon_2)^{2}
    \\
    & =  2\mu_{1} + 2 \sigma \epsilon_1.
  \end{aligned}
\end{equation}
Now we differentiate with respect to $\mu_{2}$, which gives us
\begin{equation}
  \begin{aligned}
    \partial_{\mu_{2}}f(R) & = \partial_{\mu_{2}} R^2
    \\
    & = \partial_{\mu_{2}} (z_{1}^{2} + z_{2}^{2})
    \\
    & = \partial_{\mu_{2}} (\mu_{1} + \sigma\epsilon_1)^{2} +
    \partial_{\mu_{2}}(\mu_{2} + \sigma\epsilon_2)^{2}
    \\
    & =  2\mu_{2} + 2 \sigma \epsilon_2.
  \end{aligned}
\end{equation}
Lastly, we take the partial derivative with respect to $\sigma$,
\begin{equation}
  \begin{aligned}
    \partial_{\sigma}f(R) & = \partial_{\sigma}R^2
    \\
    & = \partial_{\sigma}(z_{1}^{2} + z_{2}^{2})
    \\
    & = \partial_{\sigma}(\mu_{1} + \sigma\epsilon_1)^{2} +
    \partial_{\sigma}(\mu_{2} + \sigma\epsilon_2)^{2}
    \\
    & = 2(\mu_{1} + \sigma\epsilon_1)\epsilon_1 + 2(\mu_{2} +
    \sigma\epsilon_2)\epsilon_2
    \\
    & = 2\mu_{1}\epsilon_1 + 2 \sigma\epsilon_1^{2} +
    2\mu_{2}\epsilon_2 + 2 \sigma\epsilon_2^{2}.
  \end{aligned}
\end{equation}
Thus, our Monte Carlo estimators take the form
\begin{equation}
  \begin{aligned}
    \Gexpmuone & = \frac{1}{N}\sum_{n=1}^{N}
    \left(2\mu_{1}+2 \sigma\epsilon_1\right),
    \\
    \Gexpmutwo & = \frac{1}{N}\sum_{n=1}^{N}
    \left(2\mu_{2}+2 \sigma\epsilon_2\right),
    \\
    \Gexpsigma  & = \frac{1}{N}\sum_{n=1}^{N}
    \left(2\mu_{1}\epsilon_1 + 2 \sigma\epsilon_1^{2} +
    2\mu_{2}\epsilon_2 + 2 \sigma\epsilon_2^{2}\right).
  \end{aligned}
\end{equation}
Taking the expectation of the estimators gives us,
\begin{equation}
  \begin{aligned}
    \E[\Gexpmuone ] & = 2\mu_{1}, &
    \E[\Gexpmutwo ] & = 2\mu_{2}, &
    \E[\Gexpsigma ]  & = 4 \sigma.
  \end{aligned}
\end{equation}
The variance of the estimators is given by
\begin{equation}
  \Var[\Gexpmuone] = \Var[\Gexpmutwo] = \frac{4
  \sigma^{2}}{N}
\end{equation}
and,
\begin{equation}
  \Var[\Gexpsigma]   = \frac{4(\mu_{1}^{2} + \mu_{2}^{2}
  + 4 \sigma^{2})}{N} = \frac{4(\nu^{2} + 4\sigma^{2})}{N}.
\end{equation}
Projecting the location derivative onto $\nu$ via $\Gexpnu = (\mu_{1}/\nu)\Gexpmuone + (\mu_{2}/\nu)\Gexpmutwo$, and using the independence of the two components, gives
\begin{equation}
  \Var[\Gexpnu] = \frac{\mu_{1}^{2} + \mu_{2}^{2}}{\nu^{2}}\cdot\frac{4\sigma^{2}}{N} = \frac{4\sigma^{2}}{N},
\end{equation}
since $\nu = \norm{\vmu}$.
At $\nu = 0$ the weights $\mu_{1}/\nu$ and $\mu_{2}/\nu$ take the indeterminate form $0/0$, but the estimator extends continuously: its mean $\partial_{\nu}\E[R^{2}] = 2\nu = 0$ and its variance $4\sigma^{2}/N$ do not depend on the mean direction, so fixing any reference direction (for instance $\Gexpnu = \Gexpmuone$) yields these same values.

\textbf{Vector form.}
The component-wise derivation above can be written compactly for a general differentiable test function $f$ by differentiating through the norm $R = \norm{\vz}$.
The gradient of the norm with respect to the sample is the unit vector
\begin{equation}
  \nabla_{\vz} R = \nabla_{\vz}\norm{\vz} = \frac{\vz}{\norm{\vz}} = \hat{\vz}.
\end{equation}
Since $\vz = \vmu + \sigma\veps$ gives $\partial\vz/\partial\vmu = \mathbf{I}$ and $\partial\vz/\partial\sigma = \veps$, the chain rule yields the pathwise gradients
\begin{align}
  \nabla_{\vmu} f(R) &= f'(R)\nabla_{\vz} R = f'(R)\hat{\vz}, &
  \partial_{\sigma} f(R) &= f'(R)\hat{\vz}\cdot\veps.
\end{align}
Projecting the location derivative onto the mean direction $\hat{\vmu} = \vmu/\nu$, via $\partial_{\nu} = \hat{\vmu}\cdot\nabla_{\vmu}$, gives us the following directional derivative
\begin{equation}
  \begin{aligned}
    \partial_{\nu} f(R) &= f'(R)\hat{\vz}\cdot\hat{\vmu}\\
    &= f'(R)\cos \Phi,
  \end{aligned}
\end{equation}
where $\Phi$ is the angle between the sample $\vz$ and the mean direction $\vmu$.
Then, using the equality $\hat{\vz}\cdot\veps = \hat{\vz}\cdot(\vz - \vmu)/\sigma = (R - \nu\hat{\vz}\cdot\hat{\vmu})/\sigma$, we get that
\begin{equation}
  \partial_{\sigma} f(R) = f'(R)\frac{R - \nu\hat{\vz}\cdot\hat{\vmu}}{\sigma}.
\end{equation}
For the test function $f(R) = R^{2}$, where $f'(R) = 2R$, these reduce to the component expressions above,
\begin{align}
  \nabla_{\vmu} R^{2} &= 2R\hat{\vz} = 2\vz = 2(\vmu + \sigma\veps), &
  \partial_{\sigma} R^{2} &= 2R\hat{\vz}\cdot\veps = 2\vz\cdot\veps = 2\vmu\cdot\veps + 2\sigma\norm{\veps}^{2}.
\end{align}

\section{Derivation of the Rice implicit gradient estimator}
\label{appendix:rice_implicit}
In this section we construct an implicit pathwise estimator for the $\Rice$ distribution and the test function $f(R) = R^{2}$.
To do so, we use the CDF of the Rice distribution, denoted by $F_{R}$, which is defined as
\begin{align}
  F_{R}(r)  =  1 - Q_{1}\left(\frac{\nu}{\sigma}, \frac{r}{\sigma}\right),
\end{align}
where $Q_{1}$ is the first-order Marcum $Q$-function, which is the complementary CDF for the $\Rice$ distribution \cite{1960marcum,2005simon}.
We note that the derivatives of $Q_{1}(a,b)$ are given by
\begin{equation}
  \begin{aligned}
    \frac{\partial}{\partial a} Q_1 \left(a, b\right) & = b
    \exp\left[-\frac{a^2 + b^2}{2} \right] I_1(ab) \\
    \frac{\partial}{\partial b} Q_1 \left(a, b\right) & = -b
    \exp\left[-\frac{a^2 + b^2}{2}\right] I_0(ab),
  \end{aligned}
\end{equation}
where $I_{0}$ and $I_{1}$ are modified Bessel functions of the first kind \cite{1968pratt}.
To enable composition with a parameter-dependent map (introduced below), we rewrite the standardization function in terms of the abstract arguments of $Q_1$, defining $s: \mathbb R^2 \to \mathbb R$ as
\begin{align}
  s(a, b) = 1 - Q_1(a, b).
\end{align}
Note that $s$ coincides with $F_R$ under the substitution $a = \nu/\sigma$, $b = r/\sigma$; the two-argument form is introduced solely to make the chain rule below transparent.
The standardization function has the following Jacobian, denoted by $[\vD s(a,b)]$,
\begin{align}
  [\vD s(a,b)]  =
  \begin{bmatrix}
    -\frac{\partial}{\partial a} Q_1 \left(a, b\right) &
    -\frac{\partial}{\partial b} Q_1 \left(a, b\right)
  \end{bmatrix},
\end{align}
which we expand to
\begin{align}
  [\vD s(a,b)]  =
  \begin{bmatrix}
    -b \exp\left[-\frac{1}{2}\left(a^2 + b^2\right)\right] I_1(ab) &
    b \exp\left[-\frac{1}{2}\left(a^2 + b^2\right)\right] I_0(ab)
  \end{bmatrix}.
\end{align}
To simplify the notation, let
\begin{align}
  c(a,b) = b\exp\left[-\frac{1}{2}(a^2+b^2)\right],
\end{align}
giving us
\begin{align}
  [\vD s(a,b)] =
  \begin{bmatrix}
    -c(a,b) I_1(ab) & c(a,b) I_0(ab)
  \end{bmatrix}.
\end{align}
Now, we define the vector-valued function $\vh: \R^{3} \to \R^{2}$ as
\begin{align}
  \vh\left( R, \nu, \sigma \right) =
  \begin{bmatrix}
    \nu/\sigma \\
    R/\sigma
  \end{bmatrix}.
\end{align}
Our function $\vh$ has the Jacobian,
\begin{align}
  [\vD\vh] =
  \begin{bmatrix}
    D_{1}h_{1} & D_{2}h_{1} & D_{3}h_{1} \\
    D_{1}h_{2} & D_{2}h_{2} & D_{3}h_{2} \\
  \end{bmatrix} =
  \begin{bmatrix}
    0 & \sigma^{-1} & - \nu \sigma^{-2} \\
    \sigma^{-1} & 0 & -R \sigma^{-2} \\
  \end{bmatrix},
\end{align}
where $D_{i}$ denotes the derivative with respect to the $i$-th argument.
Using $\vh$, we can express our standardization function as a composition,
\begin{align}
  s\circ \vh = s\left(R; \nu, \sigma\right) = 1 - Q_1(\vh(R, \nu, \sigma)).
\end{align}
Using the chain rule, we get that
\begin{equation}
  \begin{aligned}
    \relax [\vD(s\circ \vh)] &= [\vD s(\vh)] [\vD \vh] \\
    & =
    \begin{bmatrix}
      D_{1} s(\vh) & D_{2} s(\vh)
    \end{bmatrix}
    \begin{bmatrix}
      0 & \sigma^{-1} & - \nu \sigma^{-2} \\
      \sigma^{-1} & 0 & -R \sigma^{-2}
    \end{bmatrix} \\
    &=
    \begin{bmatrix}
      D_{2}s(\vh) \sigma^{-1}&
      D_{1}s(\vh) \sigma^{-1} &
      - D_{1}s(\vh) \nu \sigma^{-2} - D_{2}s(\vh) R \sigma^{-2}
    \end{bmatrix}.
  \end{aligned}
\end{equation}
Substituting the components of $[\vD s]$ and writing $c = c(\nu/\sigma, R/\sigma)$, we obtain
\begin{equation}
  [\vD(s\circ \vh)] =
  \begin{bmatrix}
    \partial_R (s \circ \vh) & \partial_\nu (s \circ \vh) & \partial_\sigma (s \circ \vh)
  \end{bmatrix},
\end{equation}
where
\begin{equation}
  \begin{aligned}
    \partial_R (s \circ \vh) &= c I_0(\kappa) \sigma^{-1}, \\
    \partial_\nu (s \circ \vh) &= -c I_1(\kappa) \sigma^{-1}, \\
    \partial_\sigma (s \circ \vh) &= c \sigma^{-2}\left(I_1(\kappa)\nu - I_0(\kappa)R\right),
  \end{aligned}
\end{equation}
with
\begin{equation}
  c = c\left(\frac{\nu}{\sigma}, \frac{R}{\sigma}\right) = \frac{R}{\sigma}\exp\left[-\frac{\nu^2 + R^2}{2\sigma^2}\right], \qquad \kappa = \frac{\nu R}{\sigma^{2}}.
\end{equation}
Using the implicit reparameterization gradient formulas,
\begin{equation}
  \begin{aligned}
    \partial_\nu R &=
    -\frac{\partial_\nu (s \circ \vh)}{\partial_R (s \circ \vh)} =
    \frac{I_1(\kappa)}{I_0(\kappa)},
  \end{aligned}
\end{equation}
and
\begin{equation}
  \begin{aligned}
    \partial_\sigma R &=
    -\frac{\partial_\sigma (s \circ \vh)}{\partial_R (s \circ \vh)} =
    \sigma^{-1} \left( R - \frac{\nu I_1(\kappa)}{I_0(\kappa)}\right).
  \end{aligned}
\end{equation}
The $c$ terms cancel, leaving behind clean expressions for the gradients of the samples.

\section{French--Wilson variational implementation}
\label{app:fw_implementation}

Our French--Wilson scaling variational inference model is implemented in \prog{PyTorch} (Algorithm~\ref{alg:fw_vi}).
We fit our model using the implicit and explicit reparameterizations, using five different optimization algorithms.
For each reparameterization and optimizer pairing, we train the model $10$ times independently, each differing by the number of Monte Carlo samples, $N \in \lbrace 1,2,4, \ldots,512 \rbrace$, used for estimating the ELBO.
Every combination is replicated $10$ times (for a total of $1000$ experiments), trained for $25{,}000$ steps.

We chose the learning rates in a separate sweep, where we tested $\eta \in \{10^{-5}, 3\times10^{-5}, \dots, 10^{-1}\}$ at every $N$, using $20{,}000$ training steps for each run.
For each optimizer and estimator, we kept the rates that were stable (produced finite gradient values) at every $N$, and among those took the one with the best mean ELBO rank across $N$.
Table~\ref{tab:fw_optimizers} lists the learning rates we kept.
The two stochastic gradient descent variants were unstable with higher learning rates, and with a smaller number of Monte Carlo samples.
Thus, for the stochastic gradient descent variants we instead chose the learning rate based on stability.
Stochastic gradient descent with momentum uses a momentum of $0.9$.
Adam and AdaBelief use $(\beta_1, \beta_2) = (0.9, 0.999)$, and AdaBelief uses $\varepsilon = 10^{-16}$ with rectification and decoupled weight decay disabled.
Every remaining setting is the \prog{torch.optim} default, or the \prog{adabelief-pytorch} default for AdaBelief.

\begin{algorithm}
  \caption{French--Wilson scaling by variational inference}
  \label{alg:fw_vi}
  \begin{algorithmic}[1]
    \Require observed intensity $I_{\text{obs}}$, uncertainty $\sigma_I$, expected intensity $\Sigma$, centric flag $c$, positivity floor $\varepsilon_0$
    \Require step count $S$, Monte Carlo samples per gradient $N$, learning rate $\eta$, optimizer $\textproc{Optim}$
    \Require the explicit or the implicit reparameterization
    \Statex
    \State $a \gets \sqrt{\max(I_{\text{obs}}, 0.1)}$
    \State $b \gets 1$
    \For{$s = 1$ \textbf{to} $S$}
    \If{$c$} \Comment{centric: folded Normal, location unconstrained}
    \State $\theta \gets (\mu, \sigma) = (a,\; \mathrm{softplus}(b) + \varepsilon_0)$
    \State $q_\theta \gets \FN(\mu, \sigma)$
    \Else \Comment{acentric: Rice, location constrained positive}
    \State $\theta \gets (\nu, \sigma) = (\mathrm{softplus}(a) + \varepsilon_0,\; \mathrm{softplus}(b) + \varepsilon_0)$
    \State $q_\theta \gets \Rice(\nu, \sigma)$
    \EndIf
    \State draw $F_1, \dots, F_N \sim q_\theta$ along the explicit or implicit sampling path
    \State $\hat{\mathcal{L}} \gets \dfrac{1}{N}\displaystyle\sum_{n=1}^{N}[\log \N(I_{\text{obs}}; F_n^2, \sigma_I^2) + \log p(F_n; \Sigma, c) - \log q_\theta(F_n)]$
    \If{$\nabla_{(a,b)} \hat{\mathcal{L}}$ is finite} \Comment{explicit or implicit reparameterized gradients}
    \State $(a, b) \gets \textproc{Optim}((a,b),\, -\nabla_{(a,b)} \hat{\mathcal{L}},\, \eta)$
    \EndIf
    \EndFor
    \Statex
    \State $\hat F \gets \E_{q_\theta}[F]$, \quad $\sigma_{\hat{F}} \gets \mathrm{sd}_{q_\theta}(F)$
    \State $\hat I \gets \E_{q_\theta}[F^2]$, \quad $\sigma_{\hat I} \gets \mathrm{sd}_{q_\theta}(F^2)$
  \end{algorithmic}
\end{algorithm}

\begin{table}[h!tb]
  \centering
  \caption{
    Learning rates used for the different optimizers, one per gradient estimator.
  }
  \label{tab:fw_optimizers}
  \begin{tabular}{lcc}
    \toprule
    Optimizer & Explicit $\eta$ & Implicit $\eta$ \\
    \midrule
    SGD             & $10^{-4}$        & $10^{-4}$ \\
    SGD + momentum  & $3\times10^{-5}$ & $3\times10^{-5}$ \\
    RMSprop         & $3\times10^{-3}$ & $3\times10^{-3}$ \\
    Adam            & $3\times10^{-3}$ & $10^{-2}$ \\
    AdaBelief       & $3\times10^{-3}$ & $10^{-2}$ \\
    \bottomrule
  \end{tabular}
\end{table}

\section{Datasets and reference calculations}
\label{appendix:fw_data}

\textbf{Synthetic dataset.}
We construct an initial dataset by picking values over a grid of $I_{\text{obs}} \in \{-3, -2, \dots, 9\}$, $\sigma_I \in \{0.1, 1.1, \dots, 9.1\}$ and $\Sigma \in \{0.1, 1.1, \dots, 9.1\}$, for $1{,}300$ grid points.
Each grid point is split into centric and acentric reflections, for a total of $2{,}600$ samples.
We then use the French--Wilson scaling algorithm from the \prog{cctbx} package \cite{2002grossekunstleve} on the synthetic data.
This step rejected $360$ of the samples.
We use the remaining $2{,}240$ samples as our synthetic dataset.

\textbf{Experimental dataset.}
We use the deposited structure factors of PDB entry 9B7C, a hen egg-white lysozyme dataset in space group $P4_{3}2_{1}2$ with unit cell parameters $a = b = 79.405$\,\AA{} and $c = 37.837$\,\AA.
The \texttt{9B7C-sf.cif} file contains $45{,}196$ unique reflections spanning a resolution between $56.1$ and $1.10$\,\AA.
We use \textit{reciprocalspaceship} to read the \texttt{.cif} file, and to merge the Friedel mates.
The data is then binned into $50$ resolution shells containing equal numbers of reflections, and we use the mean intensity in each shell as the Wilson prior parameter $\Sigma$.

\textbf{Reference posteriors.}
We compute the reference posterior by quadrature, for both the synthetic dataset and the experimental dataset.
For each reflection we evaluate the amplitude posterior from equation~35 of the main text on a $20{,}001$-point grid and integrate by composite Simpson's rule.
We also report the \prog{cctbx} implementation \cite{2002grossekunstleve} of the original French--Wilson correction \cite{1978french}.

\section{Variance of the scale-derivative estimators}
\label{appendix:sigma_heatmaps}

\begin{figure}[h!tb]
  \centering
  \includegraphics[width=\textwidth]{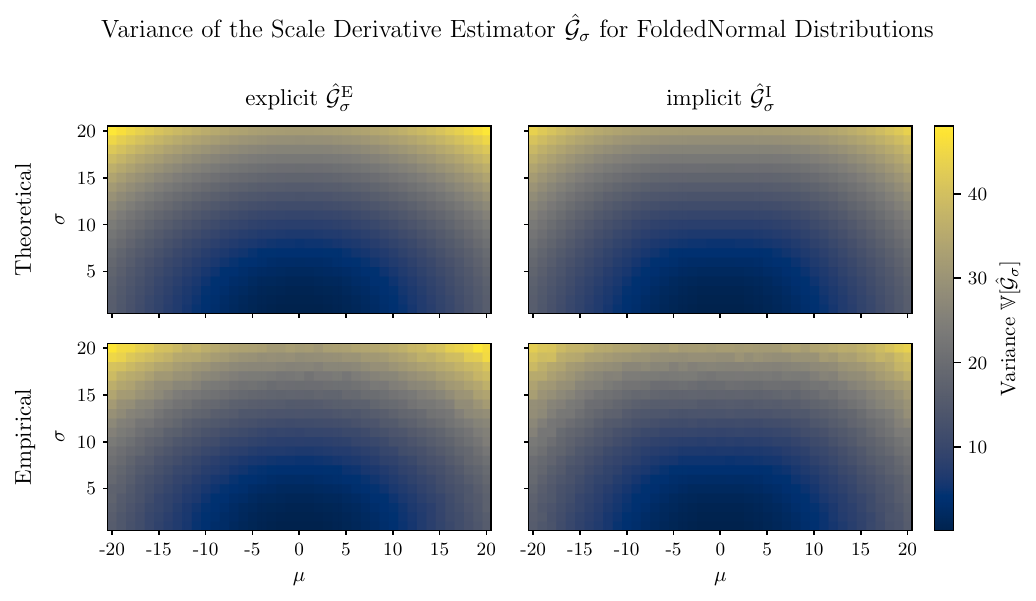}
  \caption{
    Theoretical (top) and empirical (bottom) variance of the explicit ($\Gexpsigma$) and implicit ($\Gimpsigma$) $\FN$ derivative estimators of $\partial_{\sigma}\E[Y^{2}]$ as a function of $\mu$ and $\sigma$.
    The colorbar denotes the variance value for each $(\mu,\sigma)$ pair.
    The empirical variance was computed across $10{,}000$ simulations using $100$ Monte Carlo samples each.
    The theoretical variance was calculated using equations~13 and 20 of the main text, for the explicit and implicit estimators, respectively.
  }
  \label{fig:fn_imp_exp_dfdsig_var_subplots}
\end{figure}

\begin{figure}[h!tb]
  \centering
  \includegraphics[width=\textwidth]{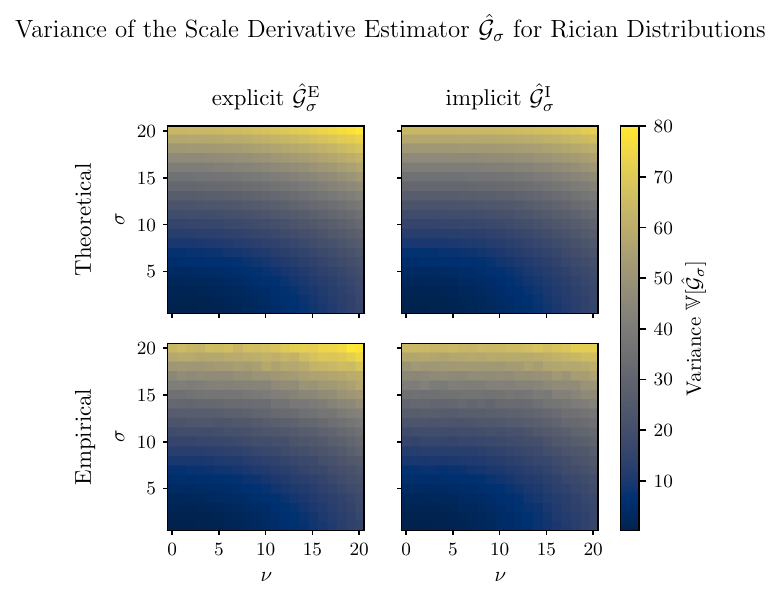}
  \caption{
    Theoretical (top) and empirical (bottom) variance of the explicit ($\Gexpsigma$) and implicit ($\Gimpsigma$) $\Rice$ derivative estimators of $\partial_{\sigma}\E[R^{2}]$ as a function of $\nu$ and $\sigma$.
    The colorbar denotes the variance value for each $(\nu,\sigma)$ pair.
    The empirical variance was computed across $10{,}000$ simulations using $100$ Monte Carlo samples each.
    The theoretical variance was calculated using equations~26 and 34 of the main text, for the explicit and implicit estimators, respectively.
  }
  \label{fig:rice_imp_exp_dfdsig_var_subplots}
\end{figure}

The variances of the explicit and implicit scale-derivative estimators across the $(\mu, \sigma)$ and $(\nu, \sigma)$ parameter grids are shown in Figs.~\ref{fig:fn_imp_exp_dfdsig_var_subplots} and~\ref{fig:rice_imp_exp_dfdsig_var_subplots}.

\section{Reference recovery under every optimization algorithm}
\label{app:fw_recovery_full}

\begin{figure}[h!tb]
  \centering
  \includegraphics[width=\textwidth]{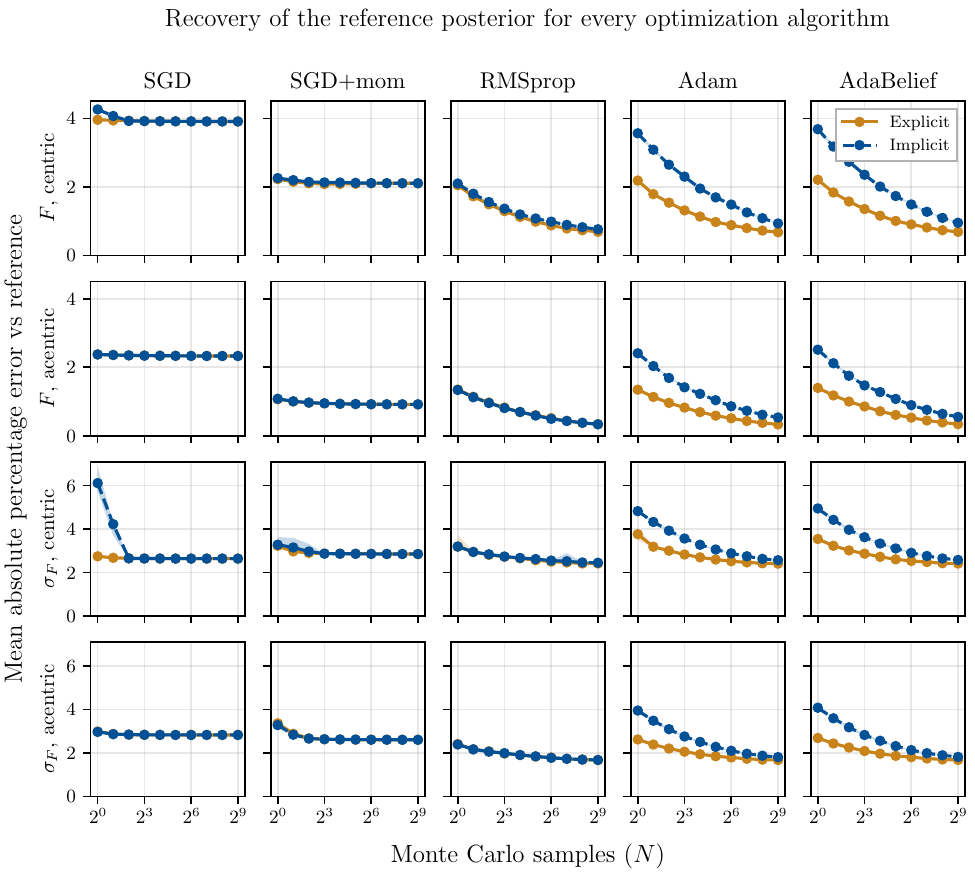}
  \caption{
    Recovery of the reference posterior as a function of the number of Monte Carlo samples $N$ used per gradient, for each of the five optimization algorithms tested.
    The first and second row show the amplitude $F$ for centric and acentric reflections, respectively.
    The third and fourth rows show the posterior width $\sigma_F$ for centric and for acentric reflections, respectively.
    The two amplitude rows share a vertical scale, and so do the two posterior-width rows.
    Lines are medians over ten replicates and bands span the interquartile range.
  }
  \label{fig:fw_recovery_full}
\end{figure}

We use the mean absolute percentage error and a correlation coefficient to score the approximate posteriors against the quadrature reference from Section~\ref{appendix:fw_data}.
The scores are reported for each optimization algorithm and reparameterization, and the scores are split by centric and acentric reflections.
Figure~\ref{fig:fw_recovery_full} plots all four errors against the number of Monte Carlo samples, and Table~\ref{tab:fw_recovery_full} gives them at $N = 512$.

\begin{table}[h!tb]
  \centering
  \caption{
    Recovery of the reference posterior at $N = 512$, for every optimization algorithm and both estimators.
    Errors are mean absolute percentage errors and correlations ($r$) are Pearson coefficients against the reference amplitude.
    Each value is the median over ten replicates.
  }
  \label{tab:fw_recovery_full}
  \begin{tabular}{llcccccc}
    \toprule
    & & \multicolumn{2}{c}{$F$ error (\%)} & \multicolumn{2}{c}{$\sigma_F$ error (\%)} & \multicolumn{2}{c}{$r$ for $F$} \\
    \cmidrule(lr){3-4}\cmidrule(lr){5-6}\cmidrule(lr){7-8}
    Algorithm & Estimator & centric & acentric & centric & acentric & centric & acentric \\
    \midrule
    SGD         & Explicit &  3.91 &  2.33 &  2.64 &  2.83 & 0.99417 & 0.99539 \\
    & Implicit &  3.91 &  2.33 &  2.64 &  2.83 & 0.99417 & 0.99539 \\
    \addlinespace
    SGD + mom.  & Explicit &  2.10 &  0.91 &  2.85 &  2.60 & 0.99896 & 0.99924 \\
    & Implicit &  2.10 &  0.91 &  2.85 &  2.60 & 0.99896 & 0.99924 \\
    \addlinespace
    RMSprop     & Explicit &  0.69 &  0.33 &  2.42 &  1.67 & 0.99975 & 0.99992 \\
    & Implicit &  0.76 &  0.33 &  2.45 &  1.68 & 0.99972 & 0.99992 \\
    \addlinespace
    Adam        & Explicit &  0.67 &  0.33 &  2.41 &  1.67 & 0.99975 & 0.99992 \\
    & Implicit &  0.93 &  0.53 &  2.57 &  1.81 & 0.99962 & 0.99979 \\
    \addlinespace
    AdaBelief   & Explicit &  0.69 &  0.34 &  2.41 &  1.68 & 0.99974 & 0.99992 \\
    & Implicit &  0.95 &  0.55 &  2.58 &  1.82 & 0.99960 & 0.99978 \\
    \bottomrule
  \end{tabular}
\end{table}

\section{Gradient variance at the initialization and at convergence}
\label{app:fw_gradvar_theta}

\begin{figure}[h!tb]
  \centering
  \includegraphics[width=\textwidth]{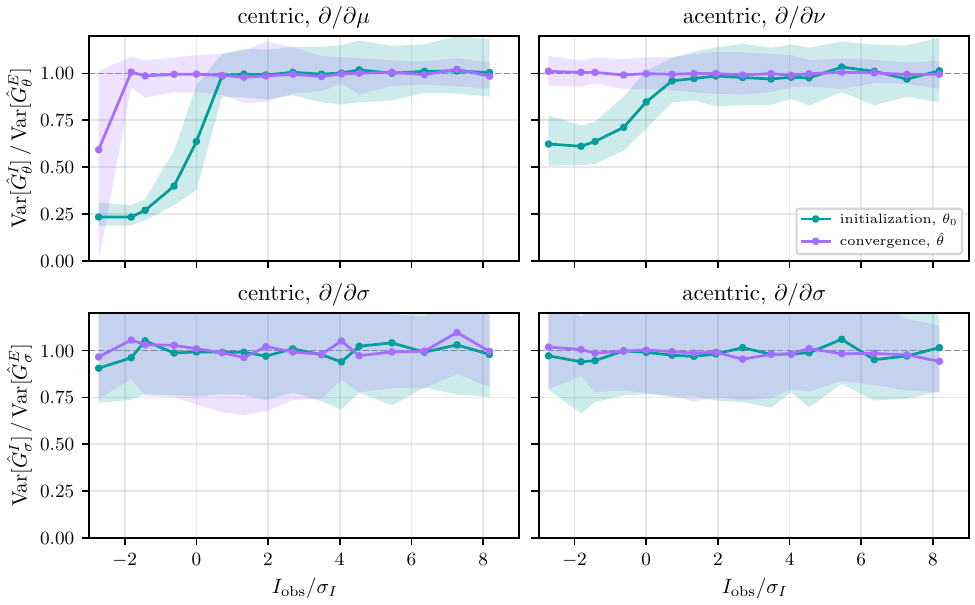}
  \caption{
    Ratio of implicit to explicit gradient variance, plotted against the observed signal-to-noise (SNR) of the reflection.
    The ratio is measured at the initialization $\theta_0$ and at the converged parameters $\hat{\theta}$ of an Adam fit after $25{,}000$ steps at $N = 512$.
    The top row shows the variance ratio for the location-derivative, and the bottom row shows the variance ratio for the scale-derivative.
    The ratio is shown separately for the centric reflections (left column) and the acentric reflections (right column).
    Each point is a median over reflections and ten replicates, and the bands span the interquartile range within each $\snr$ bin.
    The $\snr$ axis is divided into $18$ equal-width bins between $-3$ and $9$.
  }
  \label{fig:fw_gradvar_snr}
\end{figure}

\begin{figure}[h!tb]
  \centering
  \includegraphics[width=\textwidth]{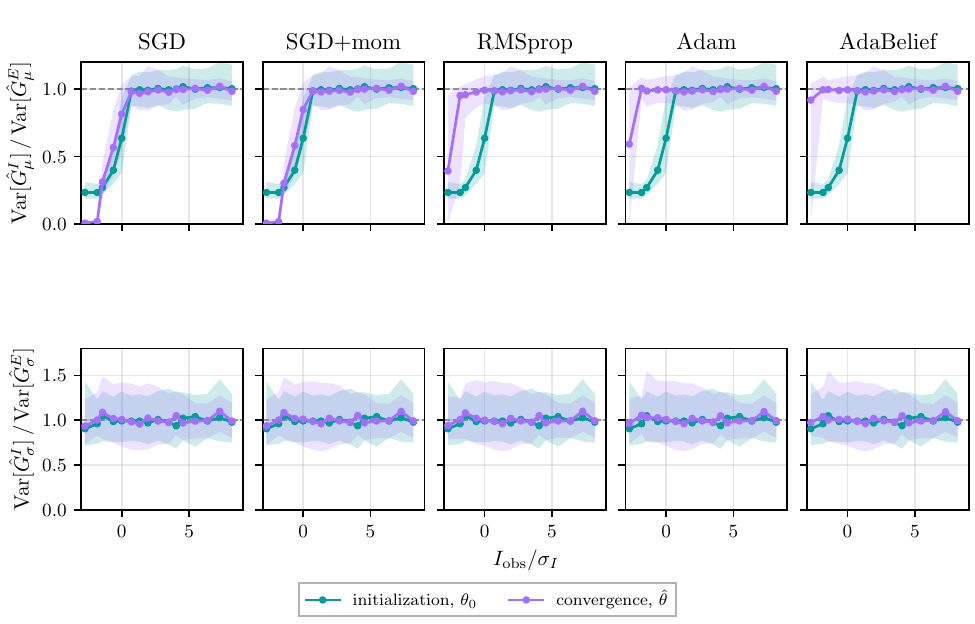}
  \caption{
    Ratio of implicit to explicit gradient variance at the initialization $\theta_0$ parameters, and at each algorithm's own converged parameters after $25{,}000$ steps at $N = 512$.
    The top row is the centric location derivative and the bottom row is the centric scale derivative.
    The initialization parameters are the same in all cases, so the $\theta_0$ curve is identical across each row.
    Points are medians over reflections and ten replicates and bands span the interquartile range within each bin.
    The SNR axis is divided into $18$ equal-width bins between $-3$ and $9$.
    The three lowest bins hold nine reflections each, and the fullest hold about three hundred.
  }
  \label{fig:fw_gradvar_theta_opt}
\end{figure}

Figure~\ref{fig:fw_gradvar_snr} shows the variance ratio at the start and end of training for the Adam optimizer; and Figure~\ref{fig:fw_gradvar_theta_opt} shows the ratio for all five optimizers we tested.
At initialization, all of the training runs are initialized at the same parameters, so the initial ratio is the same.
At convergence, each run converges to different parameters.
Thus, to measure the ratio at the end of training, we record the parameters that the implicit estimator converged to, and we use those parameters to estimate the gradient of both the explicit and implicit reparameterizations.

\section{Convergence under every optimization algorithm}
\label{app:fw_training}

\begin{figure}[h!tb]
  \centering
  \includegraphics[width=\textwidth]{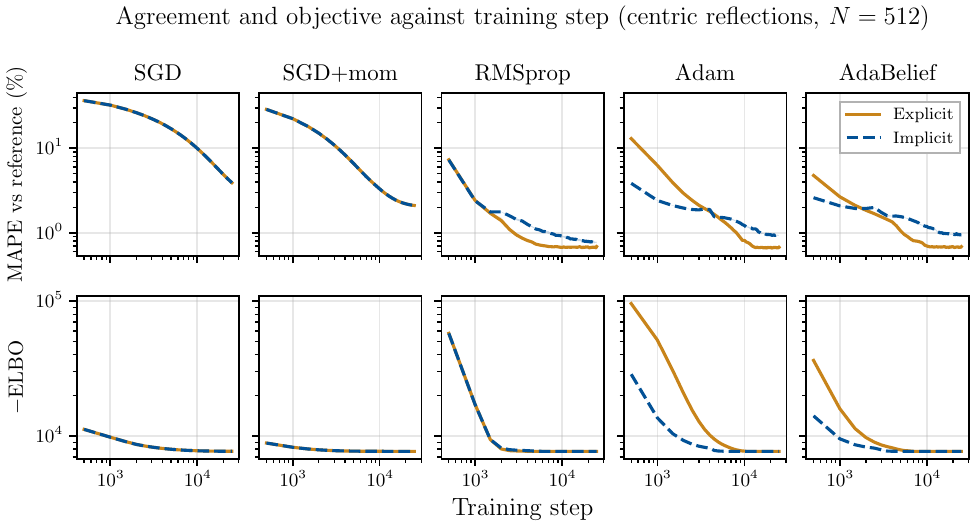}
  \caption{
    Centric amplitude error and objective against training step, one column per optimization algorithm, at $N = 512$ Monte Carlo samples per gradient.
    Top, mean absolute percentage error against the reference posterior; bottom, the negative ELBO.
    Lines are medians over ten replicates and bands span the interquartile range.
    Every algorithm drives the objective to within $0.5\%$ of the same value, so the fits agree closely on the optimum they approach.
    Stochastic gradient descent's objective is still descending at the end of training, so it has not converged.
  }
  \label{fig:fw_training}
\end{figure}

Figure~\ref{fig:fw_training} shows the centric amplitude error and the objective over training, for each of the five optimization algorithms.

\end{document}